\documentclass[]{article}
\usepackage{axodraw}
\usepackage{jheppub}

\usepackage{fancyvrb}
\usepackage{amssymb}
\usepackage{amsmath}
\usepackage{enumerate}
\usepackage{slashed}
\usepackage{graphicx}
\usepackage{color}
\usepackage[tight]{subfigure}
\usepackage{float}
\usepackage{ulem}
\usepackage{url}

\newcommand{\beq}{\begin{equation}}
\newcommand{\eeq}{\end{equation}}
\newcommand{\beqn}{\begin{eqnarray}}
\newcommand{\eeqn}{\end{eqnarray}}
\newcommand{\bmat}{\begin{pmatrix}}
\newcommand{\emat}{\end{pmatrix}}

\newcommand{\lsim}{\raisebox{-0.13cm}{~\shortstack{$<$ \\[-0.07cm] $\sim$}}~} 
\newcommand{\gsim}{\raisebox{-0.13cm}{~\shortstack{$>$ \\[-0.07cm] $\sim$}}~}

\def\gluino{{\tilde g}}
\def\squark{{\tilde q}}
\def\none{{\tilde \chi_1^0}}
\def\ntwo{{\tilde \chi_2^0}}

\def\nfour{{\tilde \chi_4^0}}
\def\chaonep{{\tilde \chi_1^{+}}}
\def\chaonem{{\tilde \chi_1^{-}}}
\def\chaone{{\tilde \chi_1^{\pm}}}
\def\chatwo{{\tilde \chi_2^{\pm}}}

\def\Fastlim{{\tt Fastlim}}

\preprint{ \begin{flushright}  CERN-TH/2014-019 \\ DESY 14-010  \\ KCL-PH-TH/2014-04  \\ MCTP-14-03 \end{flushright}}

\title{{\Fastlim}: 
a fast LHC limit calculator 
}

\author[a]{Michele Papucci}
\affiliation[a]{Michigan Center for Theoretical Physics, University of Michigan, Ann Arbor, MI 48109}
\author[b]{Kazuki Sakurai}
\affiliation[b]{Physics Department, King's College London, Strand, London WC2R 2LS, UK}
\author[c,d]{Andreas Weiler}
\affiliation[c]{CERN TH-PH Division, Meyrin, Switzerland}
\author[d]{Lisa Zeune}
\affiliation[d]{DESY, Notkestrasse 85, D-22607 Hamburg, Germany}

\abstract{
\Fastlim~is a tool to calculate conservative limits on extensions of the Standard Model from direct LHC searches without performing any Monte Carlo event generation.
The program {\it reconstructs} the visible cross sections from pre-calculated efficiency tables and cross section tables for simplified event topologies.   
As a proof of concept of the approach, we have implemented searches relevant for supersymmetric models with R-parity conservation. \Fastlim~takes the spectrum and coupling information of a given model point and provides,
for each signal region of the implemented analyses, the visible cross sections normalised to the corresponding upper limit, reported by the experiments, as well as the exclusion $p$-value. 
To demonstrate the utility of the program we study the sensitivity of the recent ATLAS missing energy searches to the parameter space of natural SUSY models.
The program structure allows the straight-forward inclusion of external efficiency tables  and
can be generalised to R-parity violating scenarios and non-SUSY models.
This paper serves as a self-contained user guide,
and indicates the conventions and approximations used.
\\
~
\\
\Fastlim~can be found at:
\href{http://cern.ch/fastlim}
{http://cern.ch/fastlim} 
}

\emailAdd{mpapucci@umich.edu}
\emailAdd{kazuki.sakurai@kcl.ac.uk} 
\emailAdd{andreas.weiler@cern.ch}
\emailAdd{lisa.zeune@desy.de}

\begin{document}
\thispagestyle{empty}

\def\thefootnote{\fnsymbol{footnote}}

\vspace{0.5cm}

\maketitle

\flushbottom
\section{Introduction}

\subsection{Motivation}

In the three years of LHC operation, ATLAS and CMS have conducted many direct new physics searches. These searches have put significant constraints on the parameter space of new physics models. The experimental collaborations have so far interpreted their results in simplified scenarios of full models like the Constrained MSSM (CMSSM)
or various simplified models, which are defined by effective Lagrangians with a small number of new physics particles and couplings, see {\it e.g.}~\cite{ArkaniHamed:2007fw,Alwall:2008ag,Alves:2011sq, Chatrchyan:2013sza}. 
On the other hand, many models have not been covered and most of the parameter space of the studied models ({\it e.g.} the MSSM with $\sim 20$ phenomenological parameters) has been left unexplored, except for a few very computationally intensive efforts in the MSSM~\cite{pmssm}.

An important question is how sensitive current analyses are to models that have so far been ignored by ATLAS and CMS and if there are holes in the coverage in the models that have been studied. Existing experimental analyses are often sensitive to alternate models, so there is not necessarily any additional effort required for the experiments in the limit setting process -- it is only a matter of reinterpreting existing results. While the experimental collaborations can do this, it is often not a good use of their computing resources and the effort required in reinterpreting results could be spent in performing new analyses. 

Recently, various groups have started to recast direct LHC searches to extract
limits on new physics scenarios, see {\it e.g.}~\cite{mass}. 
However, this usually asks for a tedious task which requires a chain of Monte Carlo (MC) simulations: event generation, detector simulation and efficiency estimation --
taking often in total a few hours to test a single model point, and a large computing cluster for days to perform parameter scans.
Tuning the MC simulations and validating the efficiency estimation for each analysis 
can also be cumbersome, especially when several analyses are considered.    

On the other hand, for models like the MSSM, the idea of Simplified Models provides the basis to decouple the (slow) MC event generation and simulation steps necessary to estimate the efficiencies, from the (much faster) limit setting steps. It is therefore desirable to develop a tool which is simple in use and can calculate a conservative limit in less than a minute per model point by using this principle. We present such a tool (\Fastlim) in this paper.
We have developed the first version of \Fastlim~specializing on R-parity conserving supersymmetric models but the approach can be generalized to any new physics model.

\subsection{The Program and Overview}

A novel feature of the program is that it does not perform any MC simulation to calculate visible cross sections.
Instead, the program {\it reconstructs} the visible cross sections from the contributions of the relevant simplified event topologies.
The visible cross section for each event topology and signal region is obtained by interpolating the pre-calculated efficiency tables and the cross section tables, which are provided together with the program.
In this approach, the reconstructed visible cross section may only be underestimated because only the available simplified topologies and searches are considered.
In other words, the limits obtained by {\tt FastLim} are always conservative.
Including additional topologies may strengthen the bounds\footnote{This approach works with most of the currently available searches which are ``cut-and-count'', but may fail with shape-analysis based searches where adding additional contributions may result in signal shapes more difficult to disentangle from the backgrounds.}. 
{\tt Fastlim} version~1.0 contains a set of event topologies which can cover the natural SUSY model parameter space.
More detailed information about version 1.0 will be given in Section~\ref{sec:version}. 
The input of the program are the masses and decay branching ratios of SUSY particles which must be given in the Supersymmetry Les Houches Accord ({\tt SLHA})~\cite{Skands:2003cj} format.
The running time is between a couple of seconds and about a half minute depending on the model point and the CPU speed.

\subsection{A Quick Start}

After the installation (for the guide, see Appendix~\ref{sec:installation}), the program can be executed by
\small\begin{verbatim}
./fastlim.py slha_files/testspectrum.slha
\end{verbatim}\normalsize
where {\tt testspectrum.slha} is a sample {\tt SLHA} spectrum file, which can be found in the {\tt slha\_files} directory.
A short summary of the results will be displayed on the screen and the output file {\tt fastlim.out} will be created.
If users want to run multiple spectrum files placed under {\tt slha\_files}, the preferred way is via the command
\small\begin{verbatim}
./ScanPoints.py slha_files/* ScanOutput
\end{verbatim}\normalsize
In this case, the output files will be created and stored in the {\tt ScanOutput} directory.

\subsection{Layout}

The rest of the paper is organised as follows:  the next section describes the method and the calculation procedure of the program. 
In Section~\ref{sec:topology}, the definition of the event topologies and our nomenclature for their identification are given.
Section~\ref{sec:output} explains the output files, in which the users can find the constraints set by the direct SUSY searches on the input model.  
Several useful approximations are introduced in Section~\ref{sec:approx}, 
which can be used to enhance the performance of the program when there is a mass degeneracy in the spectrum.
Section~\ref{sec:version} provides the detailed information on version 1.0.
In Section~\ref{sec:naturalSUSY}, we study the direct SUSY search constraints on the natural SUSY models using {\tt Fastlim}~1.0.
Section~\ref{sec:summary} is dedicated to the summary and future developments.

\section{Methodology  \label{sec:method}}

\subsection{The Traditional ``Recasting'' Approach  \label{sec:method:1}}

In a cut-and-count based analysis, experimentalists define several sets of selection cuts, called {\it signal regions}, where 
the SM events are suppressed whilst the signal events are enhanced.
One can test any SUSY model by confronting the predicted events by the theory (the sum of the SM and SUSY contributions) with the observed data in the signal regions. 
The SUSY contribution to the signal region $a$, $N_{\rm SUSY}^{(a)}$, can be written as
\beq
N_{\rm SUSY}^{(a)} = \epsilon^{(a)} \cdot \sigma_{\rm SUSY} \cdot {\cal L}_{\rm int},
\eeq
where $\epsilon^{(a)}$ is the efficiency for the signal region $a$, $\sigma_{\rm SUSY}$ is the inclusive SUSY cross section and ${\cal L}_{\rm int}$ is the integrated luminosity used in the analysis.
The efficiency and the cross section depend in general on the whole sparticle mass spectrum and couplings. 
The SUSY cross section is calculable based on the factorisation theorem and the Feynman diagram approach. Several public tools are available to calculate the total cross section beyond leading order~\cite{Beenakker:1996ed, NLLfast,nlosusy}.
One estimates the efficiency with a MC simulation, according to
\beqn
\epsilon^{(a)} = \lim_{N_{\rm MC} \to \infty} \frac{~ \#{\rm ~of~events~falling~in~signal~region~}a ~}{~ \#{\rm ~of~generated~events} ~} 
~.
\label{eq:eff}
\eeqn
There are several stages in this calculation.
First, SUSY events should be generated using event generators ({\it e.g.} {\tt Herwig}~\cite{herwig, herwigpp}, 
{\tt Pythia}~\cite{Sjostrand:2006za, pythia8} and {\tt MadGraph}~\cite{Alwall:2011uj}).
The event sample is then passed to fast detector simulation codes ({\it e.g.} {\tt Delphes}~\cite{delphes} and {\tt PGS}~\cite{pgs}) 
which should be tuned beforehand to correctly reproduce the detector response and object reconstruction criteria for a given analysis.
Finally signal region cuts must be implemented, and the efficiency is then estimated according to Eq.~(\ref{eq:eff}) using the detector level events.  

This method is generic and applicable to any model.
However, one has to tune the detector simulation and define the reconstructed objects (often on a per analysis basis),
mock-up the analyses and validate the codes in some way.
This task becomes increasingly difficult as the analyses become more elaborate and
their number and the number of signal regions increases. 
One of the solutions to this problem would be to develop a program that automatically evaluates efficiencies taking detector effects into account,
in which well validated analyses are already implemented together with the appropriate detector setups.
Along this lines, {\tt ATOM}~\cite{atom} has been developed
and already applied to some studies~\cite{Papucci:2011wy, non-deg}.\footnote{Similar programs have been put forward~\cite{Conte:2012fm,Drees:2013wra}. A framework based on the calculation of efficiencies by the experimental collaborations has been presented in~\cite{recast}.}   
{\tt ATOM} also plays a crucial role in developing {\tt Fastlim} version 1.0 as we will see in Section~\ref{sec:version}.  
Another issue is the computation time.
Even if the efficiencies were automatically calculated, the whole process, including event generation and efficiency evaluation, can easily take tens of minutes to an hour per model point. 
This becomes a crucial problem when a parameter scan is performed, requiring large computing facilities.
To overcome this problem, leveraging on the idea of simplified topologies, we take a different approach, which is described in the next subsection.

\subsection{The Method  \label{sec:method:2}}

We start by rewriting $N_{\rm SUSY}^{(a)}$.  
The SUSY contribution can be expressed as the sum of the contributions of all event topologies,
\beq
N_{\rm SUSY}^{(a)} = \sum_i^{\rm all~topologies}   \epsilon^{(a)}_i \cdot  \sigma_i \cdot {\cal L}_{\rm int},
\eeq
where $\epsilon_i^{(a)}$ is the efficiency for topology $i$, which can be calculated in the same way as in Eq.~(\ref{eq:eff}) but 
using the events with topology $i$ exclusively.
The definition of the event topologies will be illustrated in the example below and is further clarified in Section~\ref{sec:topology}. 
The cross section for topology $i$, $\sigma_i$, can be written by
the product of the production cross section and the branching ratios for the decay chains.  
The visible cross section, $\sigma_{\rm vis}^{(a)} \equiv N_{\rm SUSY}^{(a)}/{\cal L}_{\rm int}$, can be written as, for instance
\beqn
&\sigma_{\rm vis}^{(a)}& =
\nonumber \\
&& \epsilon^{(a)}_{\gluino \to q q \none:\gluino \to q q \none}(m_{\gluino}, m_{\none}) 
\cdot  \sigma_{\gluino \gluino}(m_\gluino, m_\squark) \cdot (BR_{\gluino \to qq \none})^2  +
\nonumber \\
&& \epsilon^{(a)}_{\squark \to q \none : \squark \to q \none}(m_{\squark}, m_{\none}) 
\cdot  \sigma_{\squark \squark}(m_\gluino, m_\squark) \cdot (BR_{\squark \to q \none})^2  +
\nonumber \\
&& \epsilon^{(a)}_{\gluino \to q q \none : \squark \to q \none}(m_\gluino, m_{\squark}, m_{\none}) 
\cdot  \sigma_{\gluino \squark}(m_\gluino, m_\squark) \cdot 2 \cdot BR_{\gluino \to q  q \none} \cdot BR_{\squark \to q \none} +
\nonumber \\
&& \cdots .
\label{eq:sigvis}
\eeqn
Unlike the $\epsilon^{(a)}$, the $\epsilon_i$ do not depend on all SUSY parameters 
but only on the masses and couplings of the particles appearing in the topology $i$.
Moreover, the dependence of the efficiency on the couplings is usually small~\cite{ArkaniHamed:2007fw}.  
This is because the couplings only modify angular distributions of the final state particles 
and hardly alter the hardness of the final state objects.
Current LHC searches are still inclusive enough to be not too sensitive to these effects.
In Eq.~(\ref{eq:sigvis}), the masses relevant to the efficiencies explicitly appear in the brackets.

If the decay chains in the topology $i$ are sufficiency short, the $\epsilon^{(a)}_i$ may depend only on two or three mass parameters.
For such topologies, one can pre-calculate the $\epsilon^{(a)}_i ({\bf m}_i)$ for every grid point in the parameter space, 
${\bf m}_i = \{m_i^{(1)}, m_i^{(2)}, \cdots \} $, 
and tabulate its values.
Once such tables are available, one can 
obtain the $\epsilon^{(a)}_i$ by interpolation and then reconstruct the visible cross section according to Eq.~(\ref{eq:sigvis}) without the need of carrying out a MC simulation again.
In practice, due to the ``curse of dimensionality'', it is computationally feasible to generate the efficiency tables currently only for topologies with two or three different SUSY particles\footnote{In certain cases, topologies with more than three SUSY particles may be approximated by two or three dimensional topologies, as described in Section~\ref{sec:approx}.}.
Therefore, some of the topologies may be neglected from the formula (\ref{eq:sigvis}) and in this case the reconstructed visible cross section is underestimated.
This means the derived limit is conservative.
The detailed information on the currently available efficiency tables is given in Section~\ref{sec:tables} and \ref{sec:version}. Additional tables are currently being produced and once available can be downloaded from the {\tt Fastlim} website (\href{http://cern.ch/fastlim} {http://cern.ch/fastlim}).

Similarly to the pre-calculated $\epsilon_i^{(a)}$, the program contains cross section tables for the various production modes.
The cross section is obtained by interpolating the tables during the reconstruction of the visible cross sections.
More details on the cross section calculation is given in Section~\ref{sec:tables}.

\subsection{The Calculation Procedure  \label{sec:method:3}}

\begin{figure}[t!]
\begin{center}
  \includegraphics[width=0.75\textwidth]{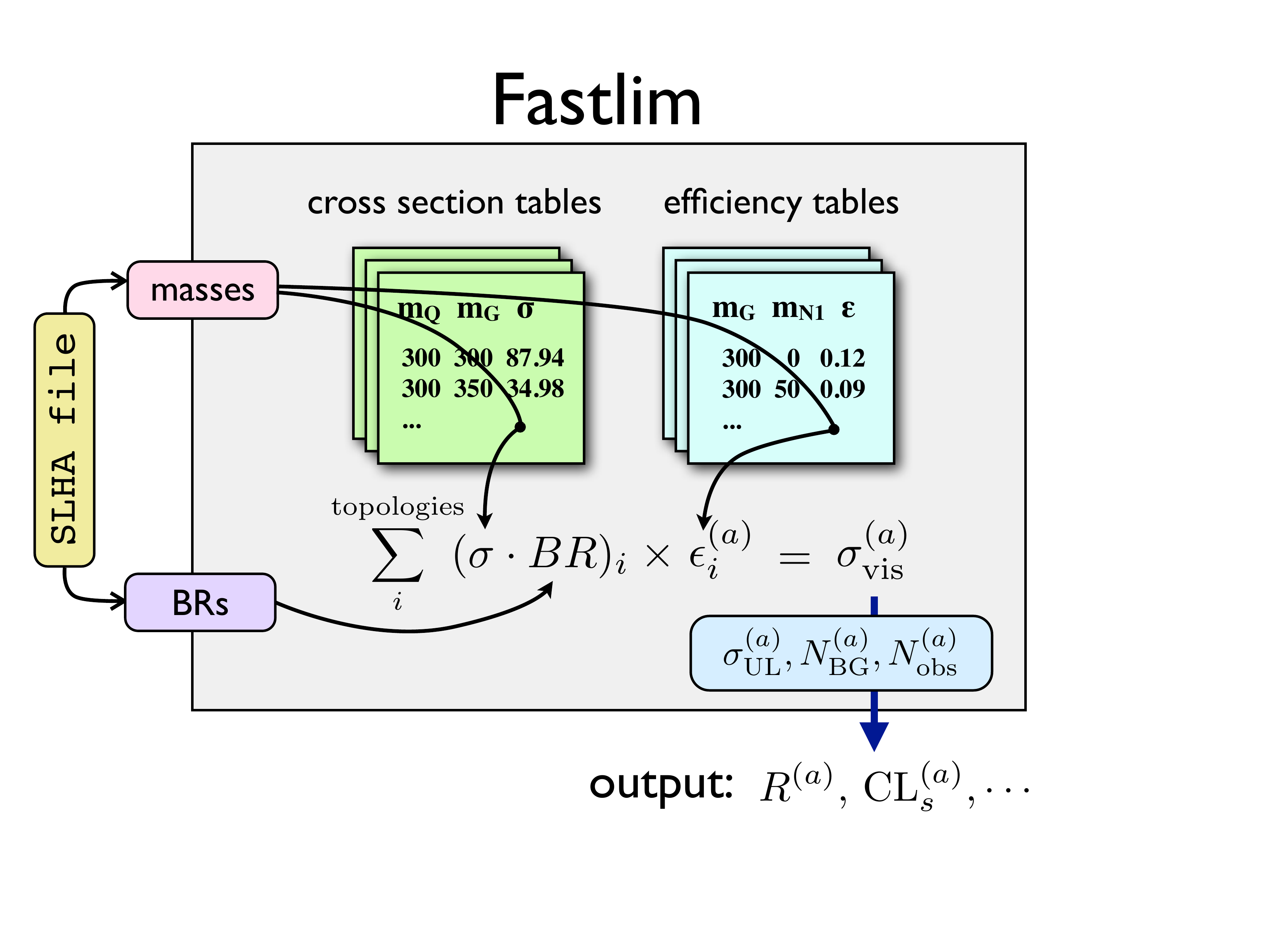}
\caption{ The structure of the program.  \label{fig:code} }
\end{center}
\end{figure}

The calculation procedure is as follows:
\begin{itemize}
\item the program first goes through all the decay chains 
starting with the SUSY particles specified in the main program file, {\tt fastlim.py}, by following the decay modes listed in the input {\tt SLHA} file.   
The program collects the branching fraction of each decay mode and calculates the total branching ratios for possible decay chains.
In this process, {\tt PySLHA}~\cite{Buckley:2013jua} is used to extract the masses and branching ratios from the {\tt SLHA} file. 
\item 
the production cross sections are then extracted for a given production mode by interpolating the cross section tables. It hen 
computes the cross sections of the event topologies, $\sigma_i$, by multiplying the production cross sections by the pairs of decay branching ratios.
The set of $\sigma_i$ contains interesting information on the model point.
The list of the cross sections for the relevant event topologies (sorted from largest to smallest) is therefore given in the output file.
\item 
a loop through all the event topologies is then performed, where the program checks for the presence of the efficiency tables for the event topology under consideration.  
If the corresponding efficiency tables are found, the efficiencies for all the signal regions are obtained by interpolating the tables.
The visible cross section for the topology, $\sigma_i^{(a)}$, is then calculated by multiplying the cross section and the efficiency. A sum over all the topologies is performed to compute the total visible cross section, $\sigma_{\rm vis}^{(a)}$, for the signal region $a$. 
The lists of $\sigma_{\rm vis}^{(a)}$ and $\sigma_i^{(a)}$ can also be found in the output file.
\item 
finally the information about the signal region $a$ necessary to set a limit is retrieved. Such information has been previously extracted from the experimental papers and it includes the 95\% CL upper limit on the visible cross section (reported by the experimental collaborations using the full likelihood), $\sigma_{\rm UL}^{(a)}$,
the contribution of the SM background, $N_{\rm BG}^{(a)}$, together with its uncertainty, 
the observed data, $N_{\rm obs}^{(a)}$, and the luminosity used for the analysis.
A convenient measure for the exclusion is the ratio between the visible cross section and its 95\% CL upper limit
$$
R^{(a)} \equiv \frac{\sigma_{\rm vis}^{(a)}}{\sigma^{(a)}_{\rm UL}} ~.
$$
The model point is excluded at the 95\% CL if $R^{(a)} > 1$.
The program may also calculate an approximate $CL_s^{(a)}$ variable by comparing $N_{\rm obs}^{(a)}$ and 
$N_{\rm BG}^{(a)} + N_{\rm SUSY}^{(a)}$ taking their uncertainties into account using an approximated likelihood $\mathcal{L} = {\rm poiss}(N_{\rm obs}^{(a)} | N_{\rm SUSY}^{(a)} + \bar b) \cdot {\rm gauss}(\bar b |  N_{\rm BG}^{(a)},\delta N_{\rm BG}^{(a)})$. 
The $CL_s^{(a)}$ variable provides a conservative exclusion criterium~\cite{Read:2002hq} since it corrects for under-fluctuations of the background. A model point is excluded if
$CL_s^{(a)} < 0.05$.  
The program outputs $R^{(a)}$ for all the signal regions and provides an approximate $CL_s^{(a)}$ if specified. An interface to {\tt RooStats}~\cite{Moneta:2010pm} is currently in testing and will be included in a future version.
\end{itemize}
A schematic diagram for the calculation procedure is shown in Fig.~\ref{fig:code}.

\section{Nomenclature of the Event Topologies\label{sec:topology}}

The guideline for a preferred definition of the event topology and its parametrisation is as follows:

\begin{itemize}

\item the event topology should be defined such that the efficiency for the topology is thoroughly determined by the masses of the on-shell SUSY particles 
appearing in the event topology when the effect of the polarisation and the spin correlation is neglected.

\item the definition and classification should be as minimal as possible, otherwise the number of event topologies becomes 
unreasonably large, requiring unnecessary efficiency tables and slowing down the computation speed.

\item the name assigned to the event topology should be as simple and intuitive as possible and must be able to identify the event topology uniquely.   
It is desirable that the name of event topologies can be directly used as a directory or file name.

\end{itemize}

Considering the first point in the guideline, the event topology should be defined by not only the final state particles 
but also the sequences of the intermediate on-shell SUSY particles in the two decay chains.
On the other hand, it does not need to specify the interactions and the off-shell particles arising in multi-body or loop-induced decays
because they only alter the decay widths and the angular distributions,
which do not have a significant impact on the efficiencies in the standard SUSY searches. 
We assume that the SUSY particles are pair produced and that 
the decay products of SUSY particles contain at most one SUSY particle at each decay vertex.
This assumption is true for R-parity conserving models, but is also realised in a large class of R-parity violating models, for which the RPV decays are present only at the end of the decay chain, due to the smallness of the RPV couplings. For those models
we allow the decay of the lightest SUSY particle (LSP) into SM particles.    
With this assumption, decay chains can be identified
by tracing the decays of SUSY particles from heavier to lighter together with 
the SM particles produced at each decay. 
It is therefore convenient to introduce a naming scheme that manifestly distinguishes R-parity even and odd particles. 
To this end, we use lower case letters for R-even particles and upper case letters for R-odd particles.
The names for R-even and R-odd particles are given in Table~\ref{tab:name}.

\begin{figure}[t!]
\begin{center}
  \includegraphics[width=0.8\textwidth]{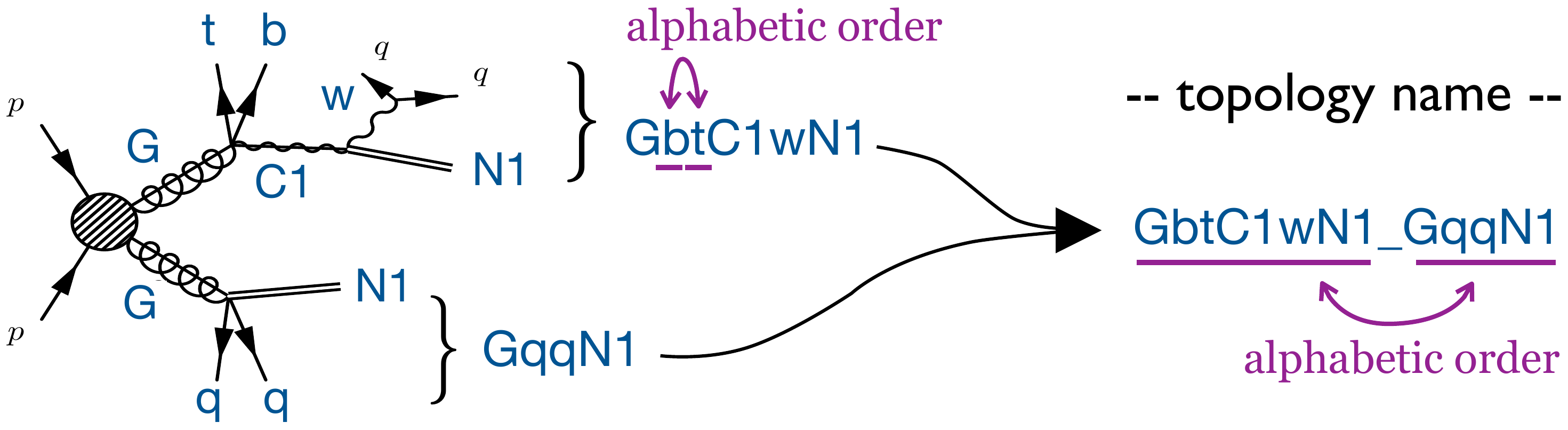}
\caption{ The naming scheme for the event topology.  \label{fig:name} }
\end{center}
\end{figure}

By using the particle names in Table~\ref{tab:name}, one can assign a unique name to each event topology
by connecting the particle names following the two decay chains.
Let us consider the event topology $pp \to \gluino \gluino$ followed by $\tilde g \to qq \none$ and $\gluino \to t b \chaone, \chaone \to W^{\pm} \none$.
We give the first decay chain the string $\tt GqqN1$.
This string is generated by joining the particle names.
In each decay, the mother SUSY particle comes first and daughter SUSY particle comes at the end, if existing.
The SM particles are placed right after their mother SUSY particle in alphabetic order.
With this rule, the string assigned to the second decay chain is uniquely determined as $\tt GbtC1wN1$.
Finally we connect the two strings in the alphabetic order and insert ``\_'' in between, which defines the name $\tt GbtC1wN1\_GqqN1$ 
for this event topology (See Fig.~\ref{fig:name}).  It is easy to realize that this prescription is unique.

\begin{table}[t!]
\begin{center}
\begin{tabular}{c || c  c  c  c  c  c  c  c | c c c c c c c}
Particle & ~$g$ & $\gamma$ & $Z$ & $h$ & $H$ & $A$ & $W^{\pm}$ & $H^{\pm}$ & $q$ & $t$ & $b$ & $e$ & $\mu$ & $\tau$ & $\nu$
\\ \hline
Name & ~\tt g & \tt gam & \tt z & \tt h & \tt h2 & \tt h3 & \tt w & \tt hp & \tt q & \tt t & \tt b & \tt e & \tt m & \tt ta & \tt n
\end{tabular}
\end{center}
\begin{center}
\begin{tabular}{c || c  c  c | c c c  c c c c}
Particle & ~$\gluino$ & $\none \cdots \nfour$ & $\chaone, \chatwo$ & $\tilde q$ & $\tilde t_1, \tilde t_2$ & $\tilde b_1, \tilde b_2$ 
& $\tilde e$ & $\tilde \mu$ & $\tilde \tau_1, \tilde \tau_2$ & $\tilde \nu, \tilde \nu_\tau$
\\ \hline 
Name & ~\tt G & ${\tt N1} \cdots {\tt N4}$ & {\tt C1}, {\tt C2} & \tt Q & {\tt T1}, {\tt T2} & \tt B1, B2
& \tt E & \tt M & \tt TAU1, TAU2 & \tt NU, NUT
\end{tabular}
\caption{The names for the R-even (top) and R-odd (bottom) particles.  R-parity is not necessarily conserved.  }
\label{tab:name}
\end{center}
\end{table}%

According to our wish list, in order to reduce the length of the decay chains,
we do not specify the decay of the SM particles    
because the decay branching ratios for the SM particles are fixed
and independent of the SUSY parameters\footnote{A possibility to account for deviations in Higgs branching ratios from the SM values may be easily accounted in future releases.}.
Similarly, we do not specify charges nor do we distinguish particles and anti-particles.
This specification is not necessary for our purpose as long as CP is conserved, since  
the branching ratio is then the same for a process and its CP conjugate.
The production cross sections are, on the other hand, different among those processes because 
the initial $pp$ state at the LHC is not CP invariant.
The ratio of the cross sections is however fixed once the masses of the produced SUSY particles are given.
Consider, for example, $pp \to \tilde d \tilde u^*$ and $pp \to \tilde d^* \tilde u$.
The productions are governed by QCD and the cross sections are fully determined by the masses of $\tilde u$ and $\tilde d$.
The ratio $\sigma(\tilde d \tilde u^*)/\sigma(\tilde d^* \tilde u)$ is therefore fixed if the masses are specified.
This means that for each grid point of the efficiency table the ratio between a process and its CP conjugation process is correctly taken into account
and is independent of the other parameters. 
Therefore, the charge of the particle does not need to be specified in the event topology for our purpose. Finally, we also do not yet distinguish between light (s)quark flavors, see however~\cite{non-deg}.

\section{The Output \label{sec:output}} 

Users can obtain information on the results at various levels of detail.
If the program is executed in the single-model-point input mode ({\it e.g.}  by {\tt ./fastlim.py slha\_files/testspectrum.slha}), 
a short summary of the results is displayed on the screen.
An example of the display output is shown in Fig.~\ref{fig:display}.
\begin{figure}[t!]
\centering
\footnotesize \begin{Verbatim}[frame=single]

----------    Cross Section    ----------
Ecm        Total   Implemented   Coverage
8TeV   750.049fb     559.215fb     74.56%

-------------------------------------------------------------------------------------------
           Analysis   E/TeV   L*fb     Signal Region:  Nev/N_UL      CLs
-------------------------------------------------------------------------------------------
ATLAS_CONF_2013_047       8   20.3           A Loose:    1.0771   0.0498   <== Exclude
ATLAS_CONF_2013_047       8   20.3          A Medium:    0.4211       --
ATLAS_CONF_2013_047       8   20.3          B Medium:    1.2380       --   <== Exclude
ATLAS_CONF_2013_047       8   20.3           B Tight:    0.0639       --
ATLAS_CONF_2013_047       8   20.3          C Medium:    4.4634       --   <== Exclude
ATLAS_CONF_2013_047       8   20.3           C Tight:    1.1229       --   <== Exclude
...

\end{Verbatim} 
\normalsize
\caption{A display output. \label{fig:display}}
\end{figure}
The first piece of information provided is how much of the total cross section is covered by
the implemented event topologies. 
If the cross section of the implemented topologies is substantially smaller than the total SUSY cross section,
the limit can be significantly underestimated.
This information is given at the beginning of the display output (See Fig.~\ref{fig:display}). 
Below the cross section information, the exclusion measures, $R^{(a)} \equiv N_{\rm SUSY}^{(a)}/N_{\rm UL}$, are given for all the signal regions.
The analysis name, the centre of mass energy, the integrated luminosity ${\cal L}_{\rm int}$ and the name of signal region are also shown in each line.
The $CL_s$ value is only displayed if $|R^{(a)} - 1 | < 0.1$ in the default setup.
If $R^{(a)} > 1$, the signal region $a$ excludes the model point at the 95\% CL.
In that case, the tag ``{\tt <== Exclude}'' appears in the end of the line of that signal region.

For more detailed information, the program also creates the output file, {\tt fastlim.out}.
The first half of an example output file is shown in Fig.~\ref{fig:output1}.
\begin{figure}[t!]
\centering
\begin{minipage}{0.75 \linewidth}
\footnotesize \begin{Verbatim}[frame=single]
##################################################
    Branching Ratio x Cross Section @ 8 TeV  
##################################################
--------------------------------------------------
Production:   Xsec/fb      Rate
     Total:   750.049   100.00%
     T1_T1:    91.441    12.19%
     B1_B1:   119.231    15.89%
       G_G:   481.097    64.14%
     T2_T2:    58.281     7.77%
--------------------------------------------------
Output processes upto 0.5%
         Process:  Br*Xsec/fb     Rate    Accum
 GbB1tN1_GbB1tN1:   238.16703   31.75%   31.75%  <== Implemented
 GbB1tN1_GtT1tN1:   177.01613   23.60%   55.35%
     B1tN1_B1tN1:   111.58518   14.88%   70.23%  <== Implemented
     T1tN1_T1tN1:    84.06936   11.21%   81.44%  <== Implemented
...

\end{Verbatim} 
\normalsize
\end{minipage}
\caption{The section dedicated to the cross section times branching ratio in the output file, {\tt fastlim.out}. \label{fig:output1}}
\end{figure}
First, the cross section for each production mode is given.
Secondly, the list of cross sections (or production cross section times branching ratios)
for the relevant event topologies is provided.  This list is sorted from the largest cross section to the smallest one.
The rate with which this process contributes to the total cross section and the accumulated rate up to the topology looked at are also shown. 
If the efficiency table for a certain event topology is implemented, the tag ``{\tt <== Implemented}'' appears.

The other half of the output is shown in Fig.~\ref{fig:output2}.
\begin{figure}[t!]
\centering
\begin{minipage}{0.75 \linewidth}
\footnotesize \begin{Verbatim}[frame=single]

############################################################
                 Analyses Details 
############################################################
------------------------------------------------------------
[ATLAS_CONF_2013_047]
0 leptons + 2-6 jets + Etmiss [squarks & gluinos] at 8TeV with $20.3fb^{-1}$
http://cds.cern.ch/record/1547563
Ecm/TeV = 8
lumi*fb = 20.3
   #----  E Medium  ----#
   Nobs:              41
   Nbg:               30.0(8.0)
   Nvis_UL[observed]: 28.6
           Process       Nev    R[obs]
             Total  189.7060    6.6277 <== Exclude
   GbB1tN1_GbB1tN1  146.4262    5.1157          
   GtT1tN1_GtT1tN1   14.5884    0.5097          
   GbB1bN1_GbB1tN1    9.9914    0.3491          
       T1tN1_T1tN1    6.3902    0.2233          
       B1tN1_B1tN1    6.2758    0.2193          
       T2bN1_T2tN1    1.9137    0.0669          
...

\end{Verbatim} 
\normalsize
\end{minipage}
\caption{The section dedicated to the information on the analyses and event topology contribution to the signal region in the output file, {\tt fastlim.out}. \label{fig:output2}}
\end{figure}
In this part the detailed information on the analysis and the constraints can be found.
The results, divided into sections, are given for each analysis.
Each section starts with the general information, providing a short description of the analysis as well as the web-link to the corresponding paper/note, 
the centre of mass energy and the integrated luminosity.
Subsequently, a summary for each signal region is presented.
It provides the name of the signal region, the number of observed events, {\tt Nobs}, the expected number of SM background events, {\tt Nbg},
and the 95\% CL upper limit on the SUSY contribution, {\tt Nvis\_UL$[$observed$]$}.
Below this information, the list of contributions of each event topology to the signal region is reported.
The event topologies are sorted in descending order from the one with the largest contribution to the smallest one. 
The contributions to the exclusion measure, {\tt R[obs]} 
(={\tt Nev/Nvis\_UL[observed]}), are also given.

\section{The Numerical Tables \label{sec:tables}}

The efficiency and cross section tables are provided in the form of a standard text file so that new tables can be added straightforwardly.
In this section, we explain the conventions for the efficiency and cross section tables.

\subsection{The Efficiency Tables} \label{sec:tab_eff}

The efficiency table file should be given for each event topology and signal region.
Two examples are shown in Fig.~\ref{fig:eff_tab}.
The header of the files describes a few remarks about the analysis and the signal region.
Below the header, each line provides the efficiency and the MC error for the SUSY masses specified at the beginning of the line
from heavier to lighter.
 The efficiency files are found for instance in
\small\begin{verbatim}
efficiency_tables/GbbN1_GbbN1/8TeV/ATLAS_CONF_2013_047/...
\end{verbatim}\normalsize
The information about the grids can be directly found in the efficiency table files.
Although the experimental collaborations have not provided their results of the signal efficiencies for the 2013 SUSY searches,
we will include them in our program whenever they will become publicly available.
The efficiency tables installed in {\tt Fastlim} 1.0 are generated by us using {\tt MadGraph 5} and {\tt ATOM}.
More detailed information is given in Section~\ref{sec:version}.

\begin{figure}
      \centering
      \begin{minipage}{0.42\linewidth}
      \begin{center}
      \footnotesize \begin{Verbatim}[frame=single]
      
   ATLAS_CONF_2013_047
   A Loose

   G     N1     Effic     Error   
   300   283    0.00117   0.00016
   300   189    0.00233   0.00024
   300   95     0.00313   0.00028
   300   1      0.00533   0.00037
   350   333    0.00149   0.00018
   350   222    0.00464   0.00033
   ...
  
      \end{Verbatim} 
      \end{center}
       \end{minipage}      
      \hspace{0.05\linewidth}       
      \begin{minipage}{0.45\linewidth}
      \begin{center}      
      \footnotesize \begin{Verbatim}[frame=single]

   ATLAS_CONF_2013_047
   A Loose

   G     T1    N1    Effic     Error     
   415   185   5     0.01148   0.00052
   415   185   1     0.01907   0.00067
   415   210   30    0.00924   0.00047
   415   210   1     0.01047   0.00050
   415   235   55    0.00779   0.00043
   415   235   28    0.00879   0.00046
   ...
  
     \end{Verbatim} 
%
%
%
      \end{center}
      \end{minipage}
              \caption{Example efficiency tables for {\tt GbbN1\_GbbN1/ATLAS\_CONF\_2013\_047/A  Loose} (left) and {\tt GtT1tN1\_GtT1tN1/ATLAS\_CONF\_2013\_047/A  Loose} (right). \label{fig:eff_tab}}      
  \end{figure}

\subsection{The Cross Section Tables \label{sec:tab_xsec}}
The cross section tables should be provided for each production mode and the centre of mass energy.
In {\tt Fastlim} 1.0, $\gluino \gluino$, $\gluino \tilde q$, $\tilde q \tilde q$ and $\tilde q \tilde q^*$ cross sections and uncertainties are generated 
by {\tt NLL\,fast}~\cite{NLLfast} combining different PDF sets, following the prescription described in Ref.~\cite{Kramer:2012bx}. 
For the stop and sbottom pair productions, the cross sections are taken from the values given by the SUSY Cross Section Working Group~\cite{SUSYxsecWG}.
The cross section table files are found for example in
\small\begin{verbatim}
xsection_tables/8TeV/NLO+NLL/...
\end{verbatim}\normalsize
or  
\small\begin{verbatim}
xsection_tables/8TeV/SUSYxsecWG/...
\end{verbatim}\normalsize

\section{The Approximations \label{sec:approx}}

\subsection{Treatment of Soft Decays  \label{sec:soft_decay}} 

Several SUSY models predict partially degenerate SUSY mass spectra.
For example, in anomaly mediation, the wino often becomes the lightest SUSY state.
Since the wino is SU(2) triplet, it leads to almost degenerate $\chaone$ and $\none$.
Another example is the higgsino LSP scenario.
In this case, two higgsino doublets have similar masses, leading to almost degenerate 
$\chaone$, $\ntwo$ and $\none$. 

If one SUSY particle decays to another which has the similar mass,   
the SM particles produced in the decay will tend to be very soft.
Such SM particles may not be observed in the detector because of the low detector acceptance and the reconstruction efficiencies.
Even if such objects are reconstructed, they hardly affect the signal region efficiency because the high-$p_T$ cuts employed in the SUSY searches
are likely to ignore such objects. 
Therefore, barring the case of dedicated analyses looking for such soft objects, if there is an event topology containing a decay associated with two nearly degenerate SUSY particles,
it may be useful to truncate the decay from the topology and redefine it as a shorter effective event topology.

Let us consider {\it e.g.} the topology {\tt GbbC1qqN1\_GbbC1qqN1}.
If the chargino, {\tt C1}, and the neutralino, {\tt N1}, are mass degenerate,
its efficiencies would be very similar to those for {\tt GbbN1\_GbbN1} because 
the light quarks from the chargino decays will be too soft to be separated from soft QCD radiation.
This observation is important because even if the efficiency tables for {\tt GbbC1qqN1\_GbbC1qqN1} are not available, 
one can nevertheless extract the efficiency from the {\tt GbbN1\_GbbN1} efficiency table, if it is implemented. 
To allow this approximation, we have implemented a {\tt Replace()} function.
In the example above the function can be used as
\small\begin{verbatim} 
Replace(procs_8, "C1qqN1", "N1"),
\end{verbatim}\normalsize 
where, {\tt procs\_8}  contains the information of all the relevant topologies together with their 8~TeV cross sections (as a {\tt Python} dictionary).
The above command replaces the string {\tt C1qqN1} by {\tt N1} in the all topologies stored in {\tt procs\_8}. 
If the event topology name generated after this truncation already exists, the contributing cross sections are 
summed: 
for the above example the cross section of {\tt GbbC1qqN1\_GbbC1qqN1} is added to the cross section of {\tt GbbN1\_GbbN1}
and the topology {\tt GbbC1qqN1\_GbbC1qqN1} is removed from {\tt procs\_8}.   
In the current version of the program such possibility is implemented by default for {\tt N1}, {\tt N2} and {\tt C1} if their mass splitting is smaller than $10\,{\rm GeV}$. The extension of such checks to other cases, via a user-defined input file is planned for the next release of {\tt Fastlim}.

Note that this replacement may introduce  topologies in which the electric charge appears to be not conserved.
For example, truncating {\tt C1qqN1} in {\tt GbbN1\_GbtC1qqN1} introduces {\tt GbbN1\_GbtN1}.
As will be discussed in subsection~\ref{sec:v_topo}, 
the program contains many such event topologies to increase the applicability to concrete models.

\subsection{Topologies with Similar Decay Structure \label{sec:similar_structure}} 

There are several event topologies among which the same efficiency table can be used.
An obvious example is {\tt T1tN1\_T1tN1} and {\tt T2tN1\_T2tN1}. In general $\tilde t_{2}$ and $\tilde t_{1}$ decay kinematics depend on their $\tilde t_{L,R}$ admixture. This is also known to accept the efficiencies of certain analyses to some level~\cite{Perelstein:2008zt, Wang:2013nwm}. While including stop polarization is a straightforward addition to {\tt Fastlim} code (which will be included in later versions), at the moment we provide efficiencies for unpolarized stops only. This allows us to present an example of another simplification feature of the {\tt Fastlim} code.

Because the polarisation effect is ignored in our calculation,
the two topologies are identical apart from the stop mass. 
As will be discussed in subsection~\ref{sec:v_topo},  
we provide the efficiency tables only for {\tt T1tN1\_T1tN1} but use them
both for {\tt T1tN1\_T1tN1} and {\tt T2tN1\_T2tN1}. 
The same efficiency tables can be also used for {\tt B1tN1\_B1tN1} and {\tt B2tN1\_B2tN1},
which may arise after truncating the soft chargino decays in {\tt B1tC1qqN1\_B1tC1qqN1} and {\tt B2tC1qqN1\_B2tC1qqN1}, respectively.

\subsection{Reduction of Multidimensional Topologies \label{sec:multi_dimensional}}  
Let us finally consider the case of {\tt GtT1tN1\_GtT2tN1}.
This event topology involves four on-shell SUSY particles: {\tt G}, {\tt T2}, {\tt T1}, {\tt N1}, and in principle requires four dimensional efficiency tables. 
However, if {\it e.g.} the masses of {\tt T1} and {\tt T2} are close to each other, one may use the efficiency tables for {\tt GtT1tN1\_GtT1tN1}, which
is three dimensional.
By default, the efficiencies for {\tt GtT1tN1\_GtT2tN1} are taken from those for {\tt GtT1tN1\_GtT1tN1}
if $(m_{\tt T2} - m_{\tt T1})/m_{\rm T2} < 0.1$.  The average mass, $(m_{\tt T2} + m_{\tt T1})/2$, is used for the mass of the intermediate particle between {\tt G} and {\tt N1} in the interpolation. This approximation can be performed automatically for particles sharing the same type of decay modes.
The same procedure and condition are used for instance for {\tt GbB1bN1\_GbB2bN1} and {\tt GbB1bN1\_GbB1bN1}. As in the case of soft decays, we plan to provide additional user control over this feature in the next {\tt Fastlim} version by suitable input configuration files.

\section{ {\tt Fastlim} version 1.0 \label{sec:version}} 

\subsection{Generation of Efficiency Tables \label{sec:v_eff}}

The simplified model efficiency tables for the 2013 SUSY searches have yet to be provided by the experimental collaborations.
The tables included in {\tt Fastlim 1.0} have therefore been pre-calculated by us using {\tt ATOM}.
The calculation procedure we used is as follows:
$5 \cdot 10^4$ events are generated using {\tt MadGraph\,5.12}~\cite{Alwall:2011uj} for each grid point in the respective SUSY mass plane (independent of the topology and the mass spectrum).
The samples include up to one extra hard parton emission at the matrix element level, matched to 
the parton shower (carried out by {\tt Pythia\,6.426}~\cite{Sjostrand:2006za})
using the MLM merging scheme~\cite{Mangano:2006rw}, where the merging scale is set to $m_{\rm SUSY}/4$ with $m_{\rm SUSY}$ being 
the mass of the heavier SUSY particles in the production.   

The event files are then passed to {\tt ATOM}~\cite{atom}, which evaluates the efficiencies for various signal regions taking  detector effects into account. 
{\tt ATOM} estimates the efficiencies for many implemented signal regions.
We have validated the implementation of the analyses in {\tt ATOM} using the cut-flow tables provided by ATLAS.
The validation results are given in Appendix~\ref{sec:validation} and the {\tt Fastlim} website (\href{http://cern.ch/fastlim} {http://cern.ch/fastlim}). 

\begin{table}[t!]
\begin{center}
\footnotesize
\begin{tabular}{c|c|c|c|c|c}
		\hline  \hline
		Name    &    Short description     &   $E_{\rm CM}$    &   ${\cal L}_{\rm int}$     &  \# SRs & Ref.      \\  \hline
                ATLAS\_CONF\_2013\_024  &  0 lepton + (2 b-)jets + MET [Heavy stop] &   8   &   20.5  &     3   &    \cite{2013_024}   \\
                ATLAS\_CONF\_2013\_035  &  3 leptons + MET [EW production]   &   8 &   20.7   &     6   &    \cite{2013_035}  \\
                ATLAS\_CONF\_2013\_037 &   1 lepton + 4(1 b-)jets + MET [Medium/heavy stop]   &   8   & 20.7  &    5   &     \cite{2013_037} \\
                ATLAS\_CONF\_2013\_047  &  0 leptons + 2-6 jets + MET [squarks \& gluinos]   &   8   &  20.3  &      10   &    \cite{2013_047} \\ 
                ATLAS\_CONF\_2013\_048  &  2 leptons (+ jets) + MET [Medium stop]   &   8   &   20.3  &     4   &    \cite{2013_048}  \\
                ATLAS\_CONF\_2013\_049  &  2 leptons + MET [EW production]   &   8  &  20.3   &      9   &    \cite{2013_049}  \\
                ATLAS\_CONF\_2013\_053  &  0 leptons + 2 b-jets + MET [Sbottom/stop]    &   8    &   20.1  &      6   &    \cite{2013_053}  \\              
                ATLAS\_CONF\_2013\_054  &  0 leptons + $\ge$ 7-10 jets + MET [squarks \& gluinos]   &   8  &  20.3   &     19   &     \cite{2013_054}  \\
                ATLAS\_CONF\_2013\_061  &  0-1 leptons + $\ge$ 3 b-jets + MET [3rd gen. squarks]   &   8 &  20.1    &     9   &   \cite{2013_061}  \\
                ATLAS\_CONF\_2013\_062   &  1-2 leptons + 3-6 jets + MET [squarks \& gluinos]   &   8   &  20.3  &    13   &    \cite{2013_062} \\ 
                ATLAS\_CONF\_2013\_093   &  1 lepton + bb(H) + Etmiss [EW production]             &   8   &  20.3  &    2   &    \cite{2013_093} \\ \hline \hline                
\end{tabular}
\caption{The analyses available in {\tt Fastlim} version~1.0. 
The units for the centre of mass energy, $E_{\rm CM}$, and the integrated luminosity, ${\cal L}_{\rm int}$, are TeV and fb$^{-1}$, respectively. 
The number of signal regions in each analysis and the references are also shown. 
\label{tab:analyses}}
\end{center}
\label{default}
\end{table}%

\subsection{The Available Analyses \label{sec:v_ana}} 

Most of the standard MET-based searches conducted by ATLAS in 2013 are available in {\tt Fastlim} version~1.0.
The list of the available analyses together with short descriptions, the centre of mass energies, the luminosities and 
the number of signal regions in the analysis are listed in Table~\ref{tab:analyses}.
The SUSY searches conducted by CMS will  be included in a future update.

\subsection{The Implemented Event Topologies \label{sec:v_topo}}

\begin{figure}[t!]
      \centering
      \begin{minipage}{0.25\linewidth}
          \begin{figure}[H]
              \includegraphics[width=0.8\linewidth]{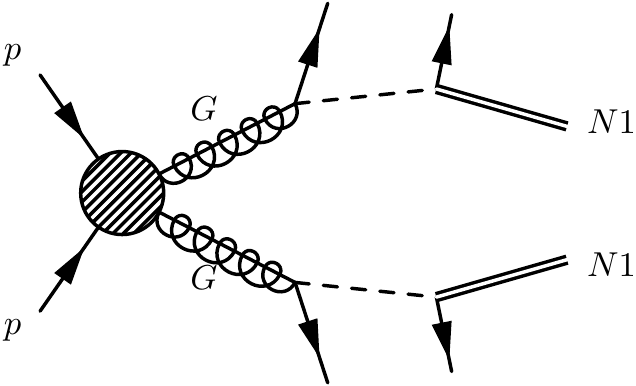}
          \end{figure}      
       %
       \begin{tabular}{c }       
        \tt GbB1bN1\_GbB1bN1 \\
        \tt GbB1bN1\_GbB1tN1 \\       
        \tt GbB1tN1\_GbB1tN1 \\               
        \tt GtT1bN1\_GtT1bN1  \\      
        \tt GtT1bN1\_GtT1tN1 \\
        \tt GtT1tN1\_GtT1tN1 \\        
       (\tt GbB2bN1\_GbB2bN1) \\
       (\tt GbB2bN1\_GbB2tN1) \\
       (\tt GbB2tN1\_GbB2tN1) \\       
       (\tt GtT2bN1\_GtT2bN1)  \\      
       (\tt GtT2bN1\_GtT2tN1) \\
       (\tt GtT2tN1\_GtT2tN1) \\       
        $[\,$\tt GbB1bN1\_GbB2bN1$\,]$ \\
        $[\,$\tt GbB1bN1\_GbB2tN1$\,]$   \\       
        $[\,$\tt GbB1tN1\_GbB2bN1$\,]$ \\
        $[\,$\tt GbB1tN1\_GbB2tN1$\,]$   \\       
        $[\,$\tt GtT1bN1\_GtT2bN1$\,]$ \\
        $[\,$\tt GtT1bN1\_GtT2tN1$\,]$   \\       
        $[\,$\tt GtT1tN1\_GtT2bN1$\,]$ \\
        $[\,$\tt GtT1tN1\_GtT2tN1$\,]$   \\       
        \end{tabular}
       \end{minipage}      
      \hspace{-0.02\linewidth}       
      \begin{minipage}{0.25\linewidth}
          \begin{figure}[H]
              \includegraphics[width=0.75\linewidth]{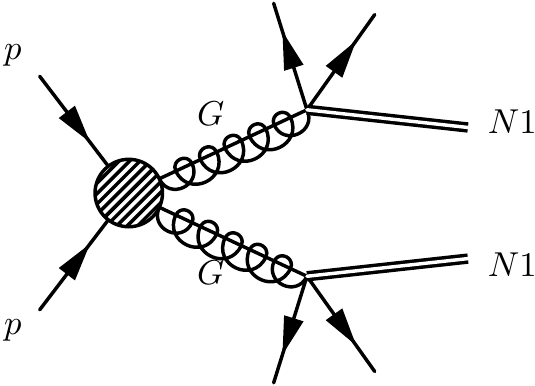}
       \end{figure}
       %
       \begin{tabular}{c}
       ~ \tt GbbN1\_GbbN1 \\
       ~ \tt GbbN1\_GbtN1 \\
       ~ \tt GbbN1\_GttN1  \\      
       ~ \tt GbbN1\_GqqN1 \\
       ~ \tt GbtN1\_GbtN1 \\
       ~ \tt GbtN1\_GttN1  \\      
       ~ \tt GbtN1\_GqqN1  \\      
       ~ \tt GttN1\_GttN1  \\      
       ~ \tt GttN1\_GqqN1  \\      
       ~ \tt GqqN1\_GqqN1  \\
                                 ~ \\
                                 ~ \\ 
                                 ~ \\                                 
                                 ~ \\ 
                                 ~ \\ 
                                 ~ \\                                                                  
                                 ~ \\
                                 ~ \\
                                 ~ \\                                 
                                 ~        
       \end{tabular}
      \end{minipage}
      \hspace{-0.02\linewidth}       
      \begin{minipage}{0.25\linewidth}
          \begin{figure}[H]
              \includegraphics[width=0.75\linewidth]{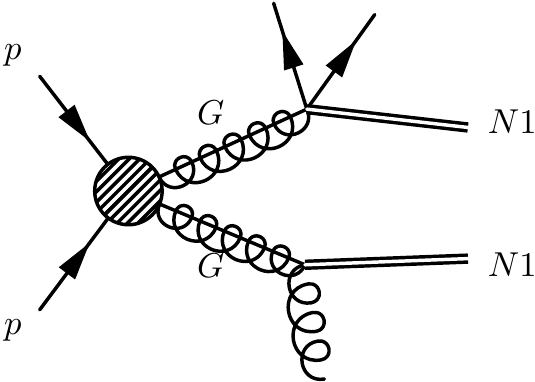}
       \end{figure}
       %
       \begin{tabular}{l }
       \tt GbbN1\_GgN1 \\
       \tt GbtN1\_GgN1 \\
       \tt GgN1\_GgN1 \\       
       \tt GgN1\_GttN1  \\      
       \tt GgN1\_GqqN1 \\
                                 ~ \\ 
                                 ~ \\
                                 ~ \\
                                 ~ \\       
                                 ~ \\ 
                                 ~ \\
                                 ~ \\ 
                                 ~ \\                                                                  
                                 ~ \\
                                 ~ \\ 
                                 ~ \\                                 
                                 ~ \\
                                 ~ \\
                                 ~ \\                                 
                                 ~
       \end{tabular}
      \end{minipage}
      \hspace{-0.02\linewidth}       
      \begin{minipage}{0.25\linewidth}
          \begin{figure}[H]
              \includegraphics[width=0.75\linewidth]{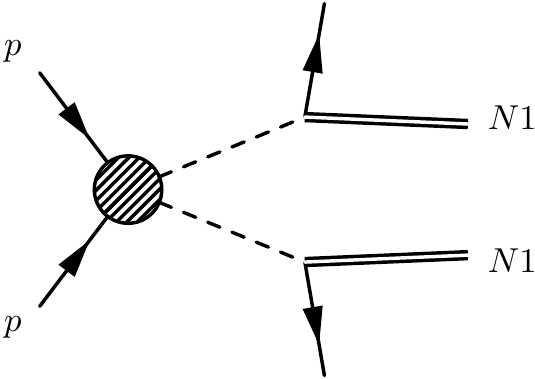}
       \end{figure}
       %
       \begin{tabular}{l }
       ~\tt T1bN1\_T1bN1 \\
       ~\tt T1bN1\_T1tN1 \\
       ~\tt T1tN1\_T1tN1 \\       
       (\tt B1bN1\_B1bN1) \\
       (\tt B1bN1\_B1tN1) \\
       (\tt B1tN1\_B1tN1) \\       
       (\tt B2bN1\_B2bN1) \\
       (\tt B2bN1\_B2tN1) \\
       (\tt B2tN1\_B2tN1) \\       
       (\tt T2bN1\_T2bN1) \\
       (\tt T2bN1\_T2tN1) \\
       (\tt T2tN1\_T2tN1) \\ 
                                 ~ \\ 
                                 ~ \\ 
                                 ~ \\                                 
                                 ~ \\ 
                                 ~ \\                                                                  
                                 ~ \\
                                 ~ \\
                                 ~              
       \end{tabular}
      \end{minipage}
              \caption{
              The event topologies whose efficiency tables are implemented in {\tt Fastlim} version~1.0.               
              The curly bracket means that the efficiencies for the topology can be taken from the efficiency tables for one of the other topologies 
              in the same group.
              On the other hand, the square bracket means that the efficiencies can be obtained only when the two intermediate SUSY masses 
              are close $m_{\tt B1} \simeq m_{\tt B2}$ or $m_{\tt T1} \simeq m_{\tt T2}$ (See subsection~\ref{sec:similar_structure} for more details.). 
              \label{fig:topo}}      
  \end{figure}

{\tt Fastlim} 1.0 contains the efficiency tables for a set of event topologies that can cover the {\it natural SUSY model} parameter space. 
By {\it natural SUSY models} we mean a type of spectra where only the gluino, left and right-handed stops, left-handed sbottom and two higgsino doublets
($\gluino$, $\tilde t_R$, $\tilde t_L$, $\tilde b_L$, $\tilde h_u$ and $\tilde h_d$) 
reside below a TeV scale and the other SUSY particles are decoupled at the LHC energy scale. 
To be more precise we list the set of event topologies implemented in {\tt Fastlim}~1.0 in Fig.~\ref{fig:topo}.
In Fig.~\ref{fig:topo},
the curly brackets mean that the efficiencies for the topology can be taken from one of the other topologies in the same group.
On the other hand, the square bracket means that the efficiencies of the event topology can be obtained only when 
the condition $m_{\tt B1} \simeq m_{\tt B2}$ or $m_{\tt T1} \simeq m_{\tt T2}$ is satisfied (See subsection~\ref{sec:similar_structure} for more details.). 

There are several event topologies in which the electric charge appears not to be conserved.
These topologies can arise after the soft decays are truncated as mentioned in subsection~\ref{sec:soft_decay}.
We also include the loop induced ${\tt G \to gN1}$ decay, which can have a sizeable branching fraction if the two-body modes and {\tt GttN1} are kinematically forbidden.  The decay rate is also enhanced if the stop and higgsino masses are small and the trilinear $A_t$ coupling is large.
These conditions can often be found in natural SUSY models.

Although the event topologies are chosen to cover natural SUSY models, many of the topologies appear also in other models.
A large rate of the gluino pair production is relatively common in a wide range of the SUSY models because of the largest colour factor of the gluino among the MSSM particles.
Many models tend to predict light stops, since the interaction between the Higgs and stops (with a large top Yukawa coupling)
pulls the stop mass down at low energies through the renormalisation group evolution, 
leading to larger branching ratios for {\tt GtT1tN1} and {\tt GttN1}.
The set of the event topologies implemented in {\tt Fastlim} 1.0 has a very good coverage also for split SUSY models if the wino or the bino is heavier than the gluino.  

Additional topologies are currently being evaluated and it will be possible to download them from the {\tt Fastlim} website (\href{http://cern.ch/fastlim} {http://cern.ch/fastlim})
as they will become available. Furthermore, any additional 3rd-party efficiency map for a topology not currently covered by {\tt Fastlim} can be easily added by formatting a text file according to the criteria exposed in Section~\ref{sec:tab_eff}. This is particularly useful to incorporate the efficiency maps that will be available from~\cite{coord}.

\section{The Constraint on Natural SUSY Models \label{sec:naturalSUSY}}

In this section, we study the direct SUSY search constraints on the natural SUSY models using {\tt Fastlim}. 
Since this is a well studied region of the SUSY parameter space~\cite{Papucci:2011wy, Hall:2011aa, Brust:2011tb, Allanach:2012vj, Kowalska:2013ica, Baer:2012uy, Berger:2012ec, Baer:2012up, Buchmueller:2013exa, Evans:2013jna},
it provides a good test case to illustrate the usage of the program.

We define natural SUSY models as a class of spectra where only gluino, left- and right-handed stops, left-handed sbottom and higgsinos 
are at energy scales accessible by the LHC.
These particles are especially sensitive to the tuning in the electroweak symmetry breaking condition 
\beq
m_Z^2 = -2( m_{H_u}^2 - |\mu|^2) + {\cal O}(\cot^2 \beta).
\label{eq:tuning}
\eeq 
This condition implies that both the higgsino mass, $\mu$, and the soft mass of the up-type Higgs, $m_{H_u}$, should not be too far from the $m_Z$ scale at the electroweak scale,
otherwise a precise cancellation is required among these parameters. 
The $m_{H_u}$ receives one-loop corrections that are proportional to 
the soft masses of the right-handed stop, $M_{U_3}$, and the third generation left-handed quark doublet, $M_{Q_3}$.
The $m_{H_u}$ also receives two-loop correction proportional to the gluino mass, $m_\gluino$.
From the naturalness point of view, we roughly expect $|\mu| \lsim M_{U_3}, M_{Q_3} \lsim m_\gluino$.
The other sparticles are not very sensitive to the fine tuning condition (\ref{eq:tuning}).
For the study below we fix the other soft masses at 3~TeV.
We calculate the sparticle spectrum and branching ratios using {\tt SUSY-HIT}~\cite{Djouadi:2006bz}. 
For the results in this section, we generated two-dimensional grids (with $\sim 500 - 1000$ points) covering slices of natural SUSY parameter space. The constraints presented below are obtained by interpolating (with {\tt Mathematica}) between the grid points. 
By using {\tt Fastlim} performing the whole study with 4836 parameter points took 18.7 hours (14 seconds per model point on average) on a single computer. 

\begin{figure}[t!]
\begin{center}
  \subfigure[]{\includegraphics[width=0.45\textwidth]{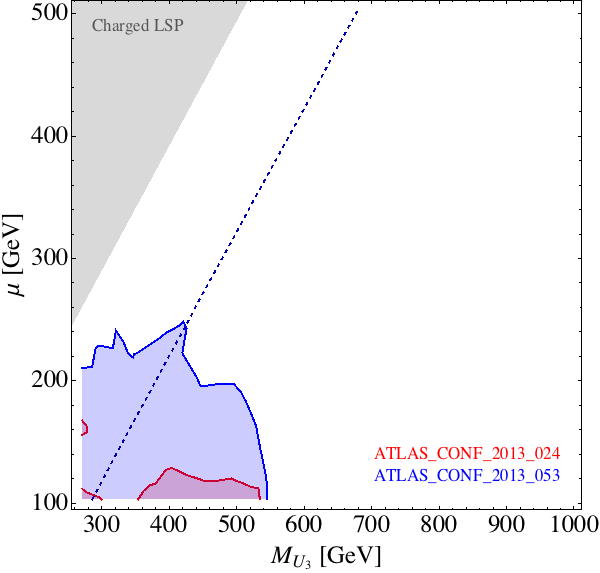}\label{fig:MU3_mu_a}} \hspace{0.2cm}
  \subfigure[]{\includegraphics[width=0.45\textwidth]{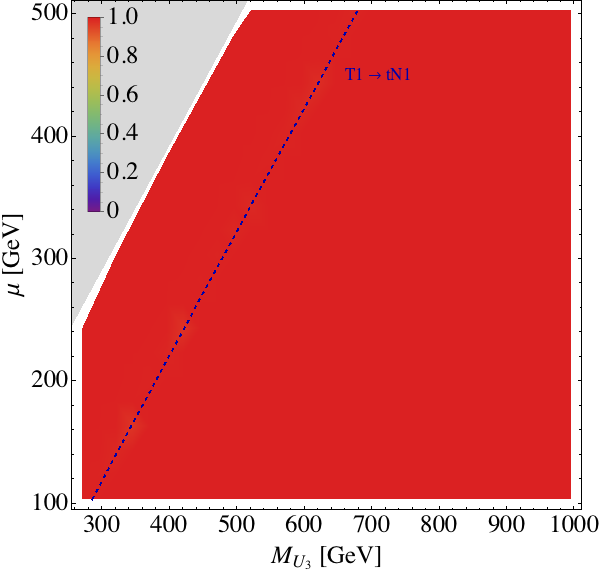}\label{fig:MU3_mu_b}} 
\caption{
Constraints from direct SUSY searches on the ($M_{U_3}$, $\mu$) plane. 
The other parameters are $m_\gluino=M_{Q_3}=M_{D_3}=3000$ GeV, $\tan \beta = 10$ and $X_t=0$.
The left plot shows the exclusion regions from the analyses listed in the plot. The right plot shows the cross section coverage, as defined in Eq.~(\ref{eq:coverage}).
The blue dashed line represents the kinematical threshold of the ${\tt T1 \to tN1}$ decay.
 \label{fig:MU3_mu} }
\end{center}
\end{figure}

In Fig.~\ref{fig:MU3_mu}, we show the direct SUSY search constraints on the ($M_{U_3}$, $\mu$) plane.
We fix the other parameters as: $M_{Q_3} = m_\gluino = 3$~TeV, $\tan \beta = 10$, $X_t  \equiv A_t - \mu \cot \beta = 0$.
Fig.~\ref{fig:MU3_mu_b} shows the cross section coverage
\beqn
{\rm Coverage} \,=\, \frac{ \sum_{i}^{\rm implemented}  \sigma_i }{ \sigma_{\rm tot} }~,
\label{eq:coverage}
\eeqn
where the numerator is the sum of the cross sections of the topologies implemented in {\tt Fastlim} 1.0.
As can be seen, {\tt Fastlim} 1.0 has a almost perfect coverage on this parameter slice.
In this model, the dominant processes are {\tt T1bN1\_T1bN1}, {\tt T1bN1\_T1tN1} and {\tt T1tN1\_T1tN1} after truncating 
the soft decays among the higgsino states: ${\tt C1, N2 \to N1}$.  
The three decays are governed by the top Yukawa coupling, but the phase space and symmetry factors give
$\sigma({\tt T1bN1\_T1tN1}) > \sigma({\tt T1bN1\_T1bN1}) > \sigma({\tt T1tN1\_T1tN1})$ in most of the parameter region.
The blue dashed line represents the kinematical limit of the ${\tt T1 \to tN1}$ decay.
The ${\tt T1bN1\_T1bN1}$ dominates in the LHS of this line. 
In the grey region, the $\tilde t_1$ becomes lighter than the $\none$ and the spectrum has a charged LSP. 
We therefore do not consider this region.

Fig.~\ref{fig:MU3_mu_a} shows the constraints from all the SUSY searches implemented in {\tt Fastlim} 1.0 (See Table~\ref{tab:analyses}).
In this plot (and the following ones of the same type)
only the names of the analyses providing an exclusion are listed on the plot, using the same colour as the exclusion contour.
The exclusion regions are plotted in descending order, starting with the top one in the list.
As can be seen, only ATLAS\_CONF\_2013\_024 and ATLAS\_CONF\_2013\_053 exclude the parameter region in the plot.
ATLAS\_CONF\_2013\_024 is designed to constrain the {\tt T1tN1\_T1tN1} topology focusing on the hadronic top decays.
Because {\tt T1tN1\_T1tN1} is subdominant in this model, the constraint from this analysis is slightly weaker than
the corresponding exclusion plot in Ref.~\cite{2013_024} assuming $Br( \tilde t_1 \to t \none) = 1$.
ATLAS\_CONF\_2013\_053, on the other hand, has been originally designed for the {\tt B1bN1\_B1bN1} topology.
In this model, {\tt T1bN1\_T1bN1} has the largest or the second largest rate among the possible topologies depending on the parameter region,
and the constraint is quite strong. 
It roughly excludes $M_{U_3} < 500$~GeV with $\mu < 200$~GeV.

\begin{figure}[t!]
\begin{center}
  \subfigure[]{\includegraphics[width=0.45\textwidth]{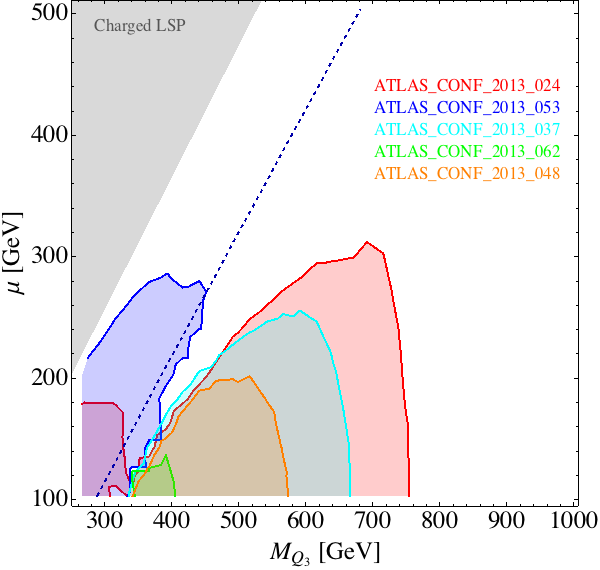}\label{fig:MQ3_mu_a}} \hspace{0.2cm}
  \subfigure[]{\includegraphics[width=0.45\textwidth]{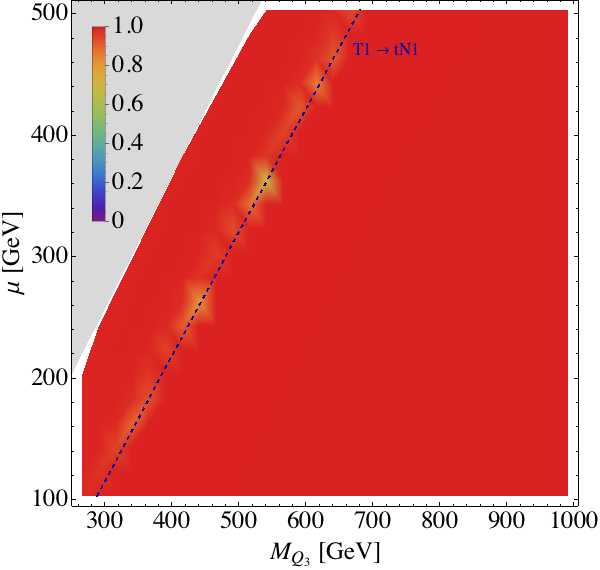}\label{fig:MQ3_mu_b}} 
\caption{ 
Constraints from direct SUSY searches on the ($M_{Q_3}$,$\mu$) plane. 
The other parameters are $m_\gluino=M_{U_3}=M_{D_3}=3000$ GeV, $\tan \beta = 10$ and $X_t=0$.  
The left plot shows the exclusion regions from the analyses listed in the plot. The right plot shows the cross section coverage, as defined in Eq.~(\ref{eq:coverage}).
The blue dashed line represents the kinematical threshold of the ${\tt T1 \to tN1}$ decay.
\label{fig:MQ3_mu} }
\end{center}
\end{figure}

Fig.~\ref{fig:MQ3_mu} shows the exclusion (left panel (a)) and the cross section coverage (right panel (b)) for the ($M_{Q_3}$, $\mu$) plane. 
The other parameters are taken as $M_{U_3} = m_\gluino = 3$~TeV, $X_t = 0$ and $\tan\beta = 10$.
The small $M_{Q_3}$ values result in both light $\tilde t_L$ and light $\tilde b_L$.
The $\tilde t_L$ is slightly heavier than the $\tilde b_L$ because of the contribution from the top quark mass 
$m_{\tilde t_L}^2 \simeq M_{Q_3}^2 + m_t^2$.
The $\tilde t_L$ and $\tilde b_L$ preferably decay to $t_R$ and $\tilde h_u$ through the interaction term 
${\cal L} \ni \epsilon^{\alpha \beta} y_t   \bar t_R (\tilde t_L, \tilde b_L)_{\alpha} (\tilde h_u^+, \tilde h_u^0)_{\beta}$. 
The ${\tt T1 \to b N1}$ and ${\tt B1 \to b N1}$ modes are instead suppressed by the bottom Yukawa coupling.
In Fig.~\ref{fig:MQ3_mu_b}, the coverage is slightly off from 100\% near the ${\tt T1 \to t N1}$ kinematical threshold line.
In this region, the three-body ${\tt T1 \to qq B1}$ decay via an off-shell $W$ boson takes a small branching fraction.
On the left hand side of the blue dashed line, ${\tt T1bN1\_T1bN1}$ and ${\tt B1bN1\_B1bN1}$ dominate.

From Fig.~\ref{fig:MQ3_mu_a}, one can see that ATLAS\_CONF\_2013\_053 only constraints the left hand side of the blue dashed line.
This can be understood because the analysis is tailored for the ${\tt T1bN1\_T1bN1}$ and ${\tt B1bN1\_B1bN1}$ topologies.
On the other side of the blue dashed line, the ${\tt T1tN1\_T1tN1}$ and ${\tt B1tN1\_B1tN1}$ topologies dominate.
In this region, ATLAS\_CONF\_2013\_024 and ATLAS\_CONF\_2013\_037 are particularly constraining 
because they are designed for the hadronic-hadronic and hadronic-leptonic top modes for  
the ${\tt T1tN1\_T1tN1}$ topology, respectively.
Overall, $M_{Q_3}$ is excluded up to 700~GeV for $\mu \lsim 250$~GeV. 
Because of the transition between different dominant decay modes,
there is a gap in the exclusion region near the blue dashed line.
In this particular region, $M_{Q_3} = 400$~GeV and $\mu = 200$~GeV is still allowed by all the analyses implemented in {\tt Fastlim}.

\begin{figure}[t]
\begin{center}
  \subfigure[]{\includegraphics[width=0.45\textwidth]{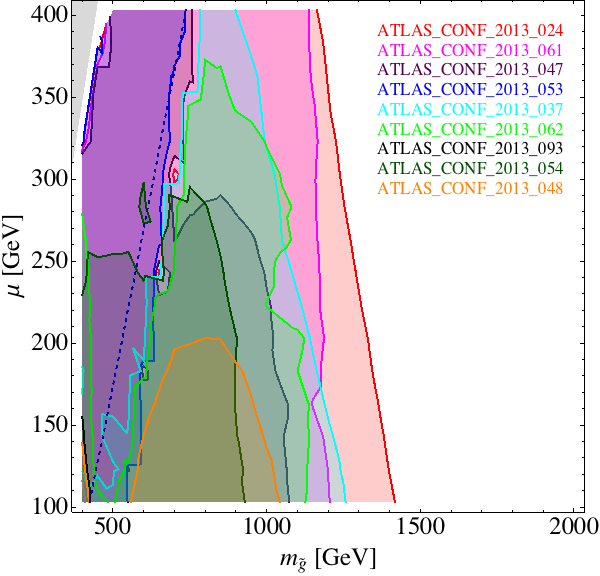}\label{fig:G_mu_a}} \hspace{0.2cm}
  \subfigure[]{\includegraphics[width=0.45\textwidth]{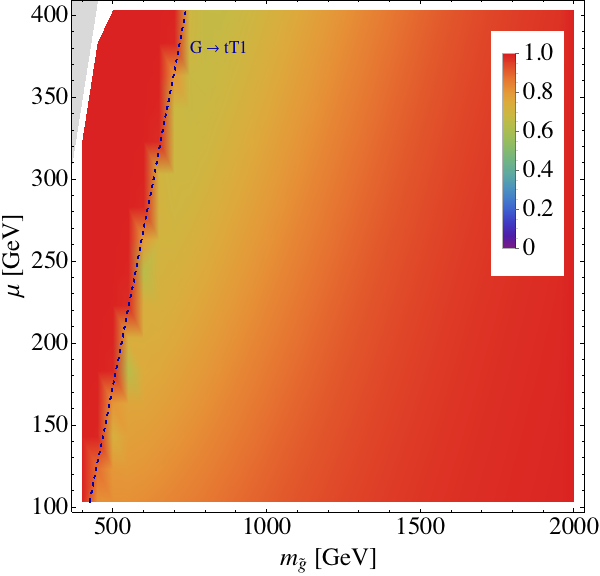}\label{fig:G_mu_b}} 
\caption{
Constraints from direct SUSY searches on the ($m_\gluino$,$\mu$) plane. 
The other parameters are $M_{U_3}=M_{D_3}=3000$ GeV, $\tan \beta = 10$ and $X_t=0$.
$M_{Q_3}$ is chosen such that the ${\tilde{t}_1}$ mass is in the middle between the $\tilde{g}$ and $\tilde{\chi}^0_1$ mass ($M_{Q_3}\simeq (m^2_{\tilde t_1}-m_t^2)^{1/2}$ with $m_{\tilde t_1} = (m_\gluino + \mu)/2$).
The left plot shows the exclusion regions from the analyses listed in the plot. The right plot shows the cross section coverage, as defined in Eq.~(\ref{eq:coverage}).
The blue dashed line represents the kinematical threshold of the ${\tt G \to tT1}$ decay.
 \label{fig:G_mu} }
\end{center}
\end{figure}

Fig.~\ref{fig:G_mu} shows the exclusion (left panel (a)) and the cross section coverage (right panel (b)) in the ($m_\gluino$, $\mu$) plane.
Here, we take $M_{U_3} = 3$~TeV, $\tan\beta = 10$, $X_t = 0$.
$M_{Q_3}$ is chosen such that the $\tilde t_1$ mass is in the middle between the $\gluino$ and $\none$ mass:
$M_{Q_3} \simeq  (m^2_{\tilde t_1} - m^2_t )^{1/2}$ with $m_{\tilde t_1} = (m_\gluino + \mu)/2$.
This condition links the stop and sbottom masses to the gluino and higgsino masses, as can be seen from the kinematical threshold for the ${\tt G \to tT1}$ decay and the charged LSP region which appears in the up left region.
Fig.~\ref{fig:G_mu_a} shows that the coverage degrades to 70\% near the ${\tt G \to t T1}$ threshold line, on its right hand side.
In this region, asymmetric gluino decays {\it e.g.}~${\tt GbB1tN1\_GtT1tN1}$ are relevant, but not implemented in {\tt Fastlim} 1.0 since they require four-dimensional grids.

Nevertheless, one can see from Fig.~\ref{fig:G_mu_a} that many analyses 
provide exclusion regions in this parameter slice because 
of the large cross section of the gluino pair production.  
Among them, ATLAS\_CONF\_2013\_024 and ATLAS\_CONF\_2013\_061 yield the most stringent constraints.  
ATLAS\_CONF\_2013\_024 mainly constrains {\tt T1tN1\_T1tN1} and {\tt B1tN1\_B1tN1} topologies,
and the bound on the gluino mass gradually decreases as the stop and sbottom masses increase together with the higgsino mass.
On the other hand, the limit from ATLAS\_CONF\_2013\_061 is almost independent of the higgsino mass.
This analysis looks for the events with 0-1 lepton plus $\geq 3$ $b$-jet, targeting the gluino pair production processes with gluino decaying to the third generation quarks either through an on- and off-shell $\tilde t_1$ and $\tilde b_1$.
The analysis roughly excludes 1.2~TeV gluino regardless of the $\mu$ parameter.

\begin{figure}[t]
\begin{center}
  \subfigure[]{\includegraphics[width=0.45\textwidth]{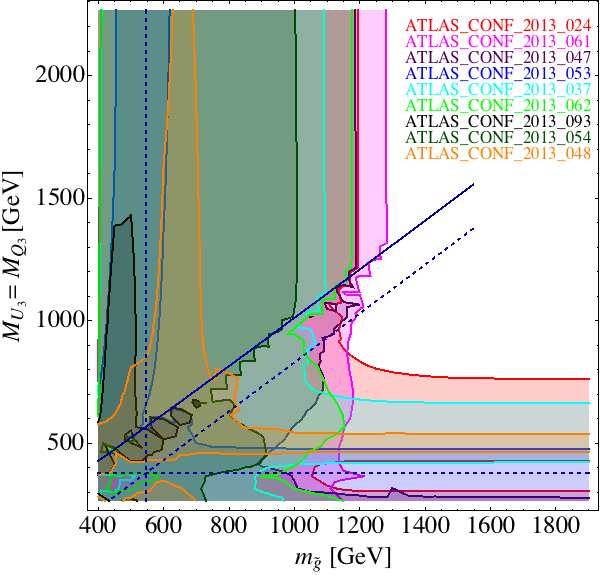}\label{fig:G_M3_a}} \hspace{0.2cm}
  \subfigure[]{\includegraphics[width=0.45\textwidth]{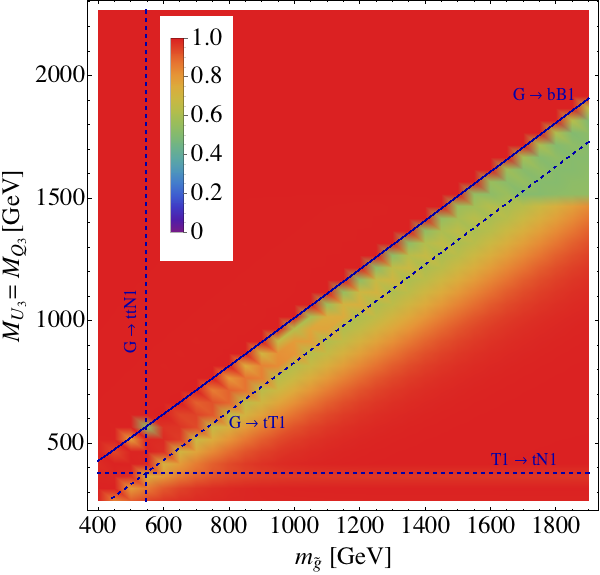}\label{fig:G_M3_b}} 
\caption{
Constraints from direct SUSY searches on the ($m_\gluino$, $M_{U_3/Q_3}$) plane. 
We set $M_{D_3}=3000$ GeV, $\tan \beta = 10,\,\mu=200$ GeV and $X_t=0$. 
The left plot shows the exclusion regions from the analyses listed in the plot. The right plot shows the cross section coverage, as defined in Eq.~(\ref{eq:coverage}).
The blue lines represent kinematical thresholds.
  \label{fig:G_M3}
  }
\end{center}
\end{figure}

We now look at the constraint on the ($m_\gluino$, $M_{U_3/Q_3}$) plane, where
we take $M_{U_3} = M_{Q_3}$, $\mu = 200$~GeV, $\tan\beta = 10$, $X_t = 0$.
Fig.~\ref{fig:G_M3_b} shows that the cross section coverage can become as small as 60\% at the vicinity of the ${\tt G \to t T1}$ threshold line.
In this region, again, the asymmetric gluino decays ({\it e.g.}~{\tt GbB1bN1\_GtT1tN1} in the region slightly above the ${\tt G \to t T1}$ threshold line,
and {\it e.g.}~{\tt GbB1bN1\_GttN1} slightly below the line) become sizeable.
One can see from Fig.~\ref{fig:G_M3_a} that the exclusions on the gluino mass and the stop mass are roughly independent of each other.
The gluino mass is excluded up to 1280~GeV, almost independently of the stop mass.
The most stringent constraint comes from ATLAS\_CONF\_2013\_061.
Near the ${\tt G \to t T1}$ threshold line the exclusion is degraded because {\tt Fastlim}~1.0 does not include the topologies with asymmetric gluino decays, 
though the degradation is only $\sim 100$~GeV on the gluino mass.
The soft mass parameters for the third generation squarks are, on the other hand, constrained up to 750~GeV.
ATLAS\_CONF\_2013\_024 provides the strongest limit in the region where $m_\gluino > 1.2$~TeV,
by excluding the stop production processes independently of the gluino mass.

\begin{figure}[t]
\begin{center}
  \subfigure[]{\includegraphics[width=0.43\textwidth]{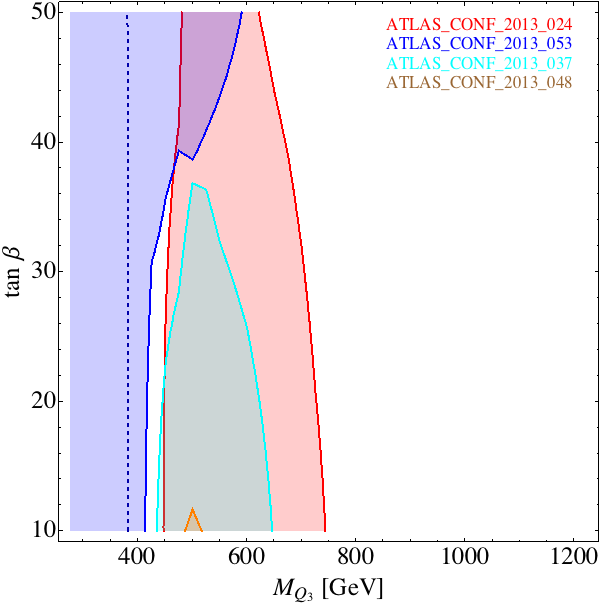}\label{fig:MQ3_TB_a}} \hspace{0.2cm}
  \subfigure[]{\includegraphics[width=0.43\textwidth]{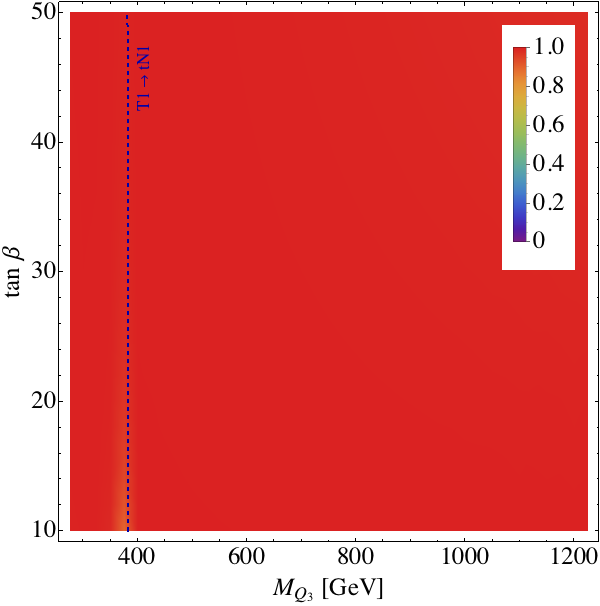}\label{fig:MQ3_TB_b}} 
\caption{
Constraints from direct SUSY searches on the ($M_{Q_3}$,$\tan \beta$) plane. 
The other parameters are $M_{D_3}=M_{U_3}=m_\gluino=3000$ GeV, $\mu=200$ GeV and $X_t=0$. 
The left plot shows the exclusion regions from the analyses listed in the plot. The right plot shows the cross section coverage, as defined in Eq.~(\ref{eq:coverage}).
The blue dashed line represents the kinematical threshold of the ${\tt T1 \to tN1}$ decay.
\label{fig:MQ3_TB} }
\end{center}
\end{figure}

In Fig.~\ref{fig:MQ3_TB}, we show the $\tan\beta$ dependence on the $M_{Q_3}$ limit.
In this parameter plane, the cross section coverage is $\sim 100\%$ across the parameter space.
The other parameters are fixed as $\mu = 200$~GeV, $X_t = 0$ and $M_{U_3} = m_\gluino = 3$~TeV.
This parameter plane intersects that of Fig.~\ref{fig:MQ3_mu_a} at $\mu = 200$~GeV, $\tan\beta = 10$.
The gap observed in Fig.~\ref{fig:MQ3_mu_a} around $M_{Q_3} \simeq 400$~GeV, $\mu = 200$~GeV 
is also seen here.
The size of $\tan\beta$ affects the branching fractions of the ${\tt T1 \to b N1}$ and ${\tt B1 \to b N1}$ modes 
since these decays are dictated by the bottom Yukawa coupling.
From $\tan\beta = 10$ to 50, $Br({\tt B1 \to b N1})$ changes from 0\% to 28\% (for $M_{Q_3} \simeq 500$ GeV). 
Because of this effect, the constraint from ATLAS\_CONF\_2013\_053 gets stronger, whilst that from ATLAS\_CONF\_2013\_024 gets
weaker as $\tan\beta$ increases.
Consequently, the gap is closed for $\tan\beta \gsim 40$.
In the large $M_{Q_3}$ region, the strongest limit comes from ATLAS\_CONF\_2013\_024 which is designed for ${\tt T1 \to t N1}$ modes.
By varying $\tan\beta$ from 10 to 50, the $M_{Q_3}$ limit changes from 750~GeV to 620~GeV.

\begin{figure}[t]
\begin{center}
  \subfigure{\includegraphics[width=0.85\textwidth]{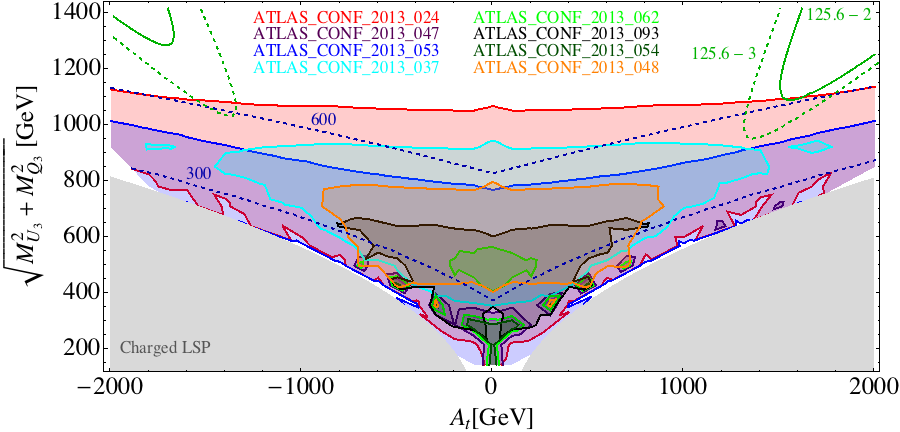}}  
    \subfigure{\includegraphics[width=0.85\textwidth]{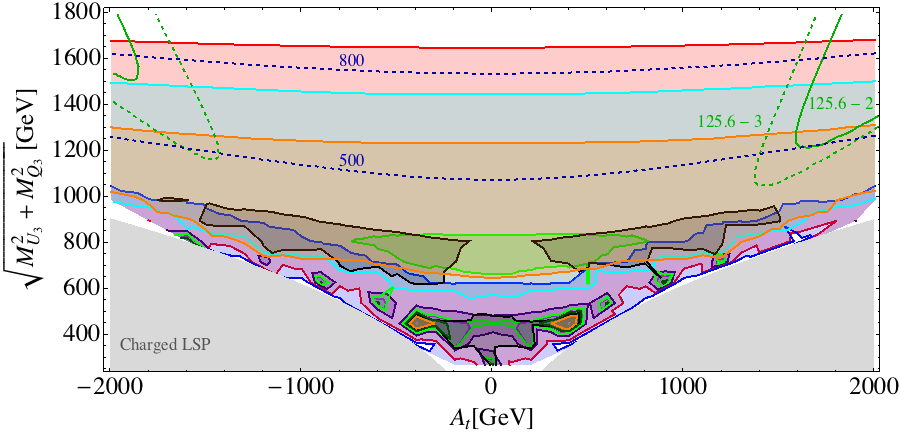}\label{fig:ExclusionMXtmuR2}}  
\caption{
Constraints from direct SUSY searches on the ($A_t$, $(M^2_{U_3} + M^2_{Q_3})^{1/2} $) plane. 
The the upper plot we choose $M_{U_3}=M_{Q_3}$ and in the lower one $M_{U_3}=2\,M_{Q_3}$.
The other parameters are $m_\gluino=M_{D_3}=3000$ GeV, $\tan \beta = 10,\,\mu=100$ GeV.
Both plots show the exclusion regions from the analyses listed in the upper plot.
The blue dashed curves show the $\tilde t_1$ mass contours.
The green curves represent the Higgs mass contours, where we allow 3 (dashed) and 2 (solid)~GeV deviation 
from the central observed value 125.6~GeV.
\label{fig:A_M3} }
\end{center}
\end{figure}

We finally show the exclusion on the ($A_t$, $(M^2_{U_3} + M^2_{Q_3})^{1/2} $) parameter plane in Fig.~\ref{fig:A_M3}.
In this plane the distance from the origin roughly corresponds to the size of the fine tuning, because the radiative correction to the up-type Higgs soft mass term is given by\footnote{This leading logarithmic approximation is generically valid for low scale SUSY breaking mediation models, while corresponding resumed expressions for high scale models can be found.}~\cite{Kitano:2006gv}
\beqn
\delta m^2_{H_u} \simeq - \frac{3 y_t^2}{8 \pi^2} \big( M_{U_3}^2 + M_{Q_3}^2 + |A_t|^2 \big) \log \Big( \frac{\rm \Lambda}{m_{\tilde t}} \Big)~,
\eeqn
where $\Lambda$ is the scale at which the SUSY breaking is mediated in the MSSM sector.
We take $M_{U_3} = M_{Q_3}$ in the upper panel, whereas $M_{U_3} = 2 \,M_{Q_3}$ in the lower panel.
The other parameters are $\mu = 100$~GeV, $\tan\beta = 10$.  

As can be seen, ATLAS\_CONF\_2013\_024 again places the most stringent limit on the soft mass for the third generation squarks
for both the $M_{U_3}/M_{Q_3} = 1$ and $  = 2$ cases.
The blue dashed curves show the $\tilde t_1$ mass contours.
One can see that the exclusion limit on $(M^2_{U_3} + M^2_{Q_3})^{1/2}$ does not change much
when $A_t$ is varied,
although the limit on the $\tilde t_1$ mass changes from 780 to 600~GeV as $|A_t|$ changes from 0 to 2~TeV (for $(M^2_{U_3}+M^2_{Q_3})^{1/2} \simeq 1$ TeV) in the $M_{U_3}/M_{Q_3} = 1$ scenario.
Increasing $|A_t|$ results in making the mass splitting between $\tilde t_1$ and $\tilde t_2$ larger.
However, the changes in the cross section times efficiency from the $\tilde t_1 \tilde t_1^*$ and $\tilde t_2 \tilde t_2^*$ processes 
tend to cancel each other and the resulting visible cross sections are more or less stable against the variation of $|A_t|$.
For $M_{U_3}/M_{Q_3} = 2$ scenario, $\tilde t_1$ is mostly composed of $\tilde t_L$ and 
the dependence of $|A_t|$ on the $\tilde t_1$ mass itself is very mild.

The green curves represent the Higgs mass contours, where we allow 3 (dashed) and 2 (solid)~GeV deviation 
from the central observed value, taking the theory uncertainties into account.
We have calculated the Higgs mass using {\tt FeynHiggs~2.9.4}~\cite{Frank:2006yh}.
Most of the parameter space is constrained by the Higgs mass measurement in the $M_{U_3}/M_{Q_3} = 1$ scenario,
whereas in the $M_{U_3}/M_{Q_3} = 2$ scenario the LHC constraints from the 8~TeV data exclude a significant part of the parameter space where the Higgs mass condition is satisfied.

\section{Discussion and Future Developments\label{sec:summary}}

In this paper we presented a program ({\tt Fastlim}) which calculates the constraints from direct SUSY collider searches starting from a given {\tt SLHA} model input file.
A novel feature of the program is that it does not run any MC simulation to calculate the visible cross section.
The program instead {\it reconstructs} the visible cross section for each signal region 
by adding the contributions from various event topologies.
The cross section and efficiencies for each event topology and each search signal region are obtained by interpolating the pre-calculated cross section and efficiency tables.
Similar ideas have also been discussed in the literature~\cite{Alves:2011sq, Gutschow:2012pw, ArkaniHamed:2007fw}.

A similar but different approach has recently been taken and implemented in~\cite{Kraml:2013mwa}.
In this approach, one checks if the model contains the event topologies on which the cross section upper limit is reported 
by the experimental collaborations.
If such event topologies are found, the program calculates the cross section time branching ratios for those topologies
and if one of them exceeds the experimental upper limit, it declares the model is excluded.
This method provides generally weaker (but more conservative) limits compared to our approach since the exclusion is made essentially from a single event topology and no recasting for topologies not covered by the experiments is performed. 

To implement our visible cross section reconstruction method, we have introduced a minimal and intuitive naming scheme for the event topology,
which can also be conveniently used as a directory or file name for the efficiency tables.  
We have also introduced useful approximations which are used to enhance the applicability and speed of the program. Such approximations include shortening the decay chains in presence of mass degeneracies in the spectrum, or recycling efficiency maps in presence of different SUSY particles sharing similar decay modes.

To demonstrate the utility of the program, we have studied the direct SUSY search constraints on  natural SUSY models.
Using the results of the 2013 ATLAS SUSY searches, we have found that the stop is excluded up to about $700$~GeV with $\mu \lsim 200$~GeV,
whereas the gluino mass is excluded up to about $1.2$~TeV with $\mu \lsim 400$~GeV. 
When $A_t$ is varied, we found that the direct SUSY search constraint can be more stringent compared to the Higgs mass constraint 
in some parameter region, which was not the case when the 7~TeV data was considered~\cite{Papucci:2011wy}. 
Running {\tt Fastlim} to extract the limits on the 4836 parameter points composing the two-dimensional plots shown in this paper took 18.7 hours (14 seconds per point on average) on a single computer.

{\tt Fastlim} version 1.0 contains the set of event topologies shown in Fig.~\ref{fig:topo}.
These topologies cover the natural SUSY model parameter space very well but
they can also cover other models such as split SUSY models with a decoupled wino or bino.
More topologies and analyses will be implemented in future updates very soon, thus extending the range of applicability of the approach.
The code structure is flexible and the efficiency tables provided from other collaborations can be included straightforwardly.
We particularly hope that the experimental collaborations will directly provide their efficiencies in a table format so that
the results can be included and thus reinterpreted in a wide range of the SUSY models.
Recasting LHC analyses to extend the number of topologies covered is becoming a coordinated effort~\cite{coord}.
Once enough topologies will be available {\tt Fastlim} can be used for computationally lean pMSSM studies,
which may give new insights into interesting SUSY models based on the LHC data.

\acknowledgments

We want to thank
S.~Caron,
T.~Cohen,
M.~D'Onofrio,
B.~Fuks,
E.~Halkiadakis,
S.~Heinemeyer,
A.~Hoeker,
K.~Howe,
M.~Mangano,
Z.~Marshall,
M.~Pierini,
S.~Pl{\"a}tzer,
S.~Prestel,
M.~Tonini,
J.~Wacker,
G.~Weiglein, and
F.~Wuerthwein for helpful discussions.
This work has partially been supported by 
the Collaborative Research Center SFB676 of the DFG,
``Particles, Strings and the early Universe".
The work of K.S. was supported in part by
the London Centre for Terauniverse Studies (LCTS), using funding from
the European Research Council 
via the Advanced Investigator Grant 267352.
K.S. thanks the CERN Theory Group for hospitality during part of this work.
K.S. thanks T.~Becker, L.~Oppermann, V.~Selk, M.~Wulf for helpful discussions.

\appendix

\section{Installation \label{sec:installation}}

To run \Fastlim~on your system, first download the latest version of the program via:
\begin{flushleft}
\href{http://cern.ch/fastlim}
{http://cern.ch/fastlim} 
\end{flushleft}
{\tt Fastlim} is based on the following software:
\begin{itemize}
\item {\tt Python}~\cite{python}, typically preinstalled
\item {\tt NumPy}~\cite{numpy} and {\tt SciPy}~\cite{scipy}, whose installation is recommended
\end{itemize}
{\tt Fastlim} was developed using {\tt Python} version 2.7, {\tt NumPy} version 1.7.1
and {\tt SciPy} version 0.12.0~\footnote{The compatibility of {\tt Fastlim} with different versions has been tested in cases.  {\tt Fastlim}  can be used also with {\tt Python} version 2.6, but the current version of our code is incompatible with {\tt Python} version 3. {\tt NumPy} versions newer than 1.6.1 and {\tt SciPy} versions newer than 0.10.0 should work.}.
The default interpolation routine in {\tt Fastlim} uses {\tt NumPy}
and {\tt SciPy}. If these packages 
are not available, 
{\tt Fastlim} switches to a cruder nearest-neighbor interpolation.
The $CL_s$ calculation relies on {\tt Numpy}.
If the user is only interested in the $R^{(a)}$ values, it is possible to use {\tt Fastlim} without {\tt NumPy}/{\tt SciPy},
however we strongly recommend to install these packages.
Details on the installation of {\tt NumPy}/{\tt SciPy} can be found on the {\tt Fastlim} website.
After downloading, run the commands
\small\begin{verbatim}
tar zxvf fastlim-*.*
cd fastlim-*.*
\end{verbatim}\normalsize
to extract the tarball 
and enter the directory.
No further installation is required.

Bugs and feature requests may be reported by sending an email to \href{mailto:fastlim.developers@gmail.com}{fastlim.developers@gmail.com}.

\newpage
\section{Validation \label{sec:validation}}

The efficiency tables installed in {\tt Fastlim} version~1.0 are generated by {\tt ATOM}~\cite{atom}, in which we have implemented various 2013 ATLAS analyses.
We have validated our implementation mostly using the cut-flow tables provided by ATLAS. 
For ATLAS\_CONF\_2013\_062, the truth-level information is used in the ATLAS cut-flow tables, which prevents us from
comparing our efficiencies and ATLAS's.  
We validated this analysis among the collaboration by cross checking the two independent implementation of the analysis.    
For ATLAS\_CONF\_2013\_053 and ATLAS\_CONF\_2013\_054 the cut-flow tables are not provided.
We therefore validated them using the simplified model exclusion plots given in the manuscripts~\cite{2013_053, 2013_054}.
The discrepancies between ATLAS and {\tt ATOM} are within $10-20\%$ for most of the signal regions. 
For the worst signal region the disagreement is about $30\%$.
Such deviations often come from jet veto cuts, which possess large theoretical uncertainties.  
Further tuning of the {\tt ATOM} detector response may improve the situation. Updated grids and validation tables will be provided in future {\tt Fastlim} versions and on the website (\href{http://cern.ch/fastlim}{http://cern.ch/fastlim}).

In what follows, we present a normalised efficiency for each stage of the cut in the cut-flow tables.
ATLAS sometimes calculates the efficiency after the trigger requirement, whereas we do it before that.
For such cases, the comparison is only reasonable after the cut to which the trigger requirement is subjected.     
The efficiency is therefore normalised to the efficiency of such a cut, which appears the first in the table.
In the tables, we use the following variables:
\beqn
\epsilon_{\rm ATLAS/ATOM}:&& {\rm normalised ~efficiency ~by ~ATLAS/ATOM},  \nonumber \\
R_{\rm ATLAS/ATOM}:&&  {\rm the~ efficiency ~ratio ~against ~the~efficiency~of~the~cut~one ~before},  \nonumber \\
{\rm Stat}:&&  {\rm the~ Monte ~Carlo ~uncertainty ~for ~the ~ATOM ~efficiency~or~efficiency ~ratio}.  \nonumber 
\eeqn


\subsection*{ATLAS\_CONF\_2013\_024}

\begin{itemize}
\item The events are generated using {\tt Herwig++~2.5.2}~\cite{herwigpp} throughout this analysis. 
\end{itemize}

\begin{figure}[h!]
\begin{center}
  \includegraphics[width=1.0\textwidth]{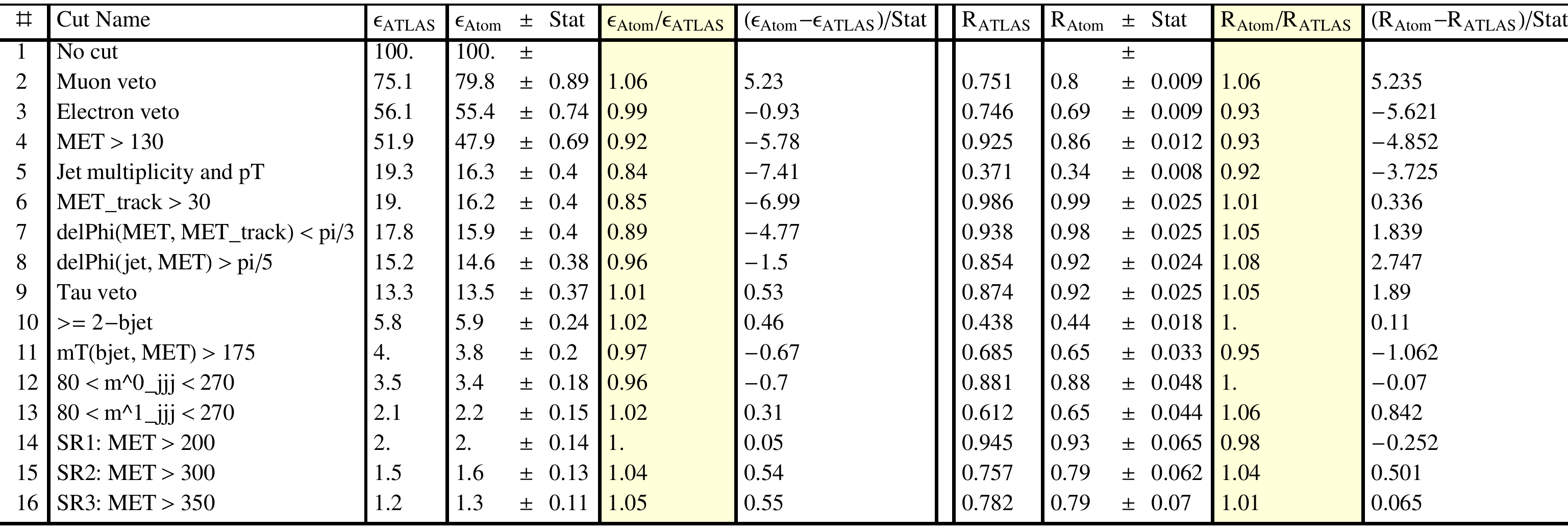}
    \vspace{-5mm}
\caption{ The efficiencies in the cut-flow for ATLAS\_CONF\_2013\_024.
$10^4$ events of $pp \to \tilde t_R \tilde t_R^* \to t \tilde \chi_1^0 \bar t \tilde \chi_1^0$ process are used.
The stop and neutralino masses are 600~GeV and 0~GeV, respectively. 
\label{fig:v_2013_024_tR} 
}
\end{center}
\end{figure}

\begin{figure}[h!]
\begin{center}
  \includegraphics[width=1.0\textwidth]{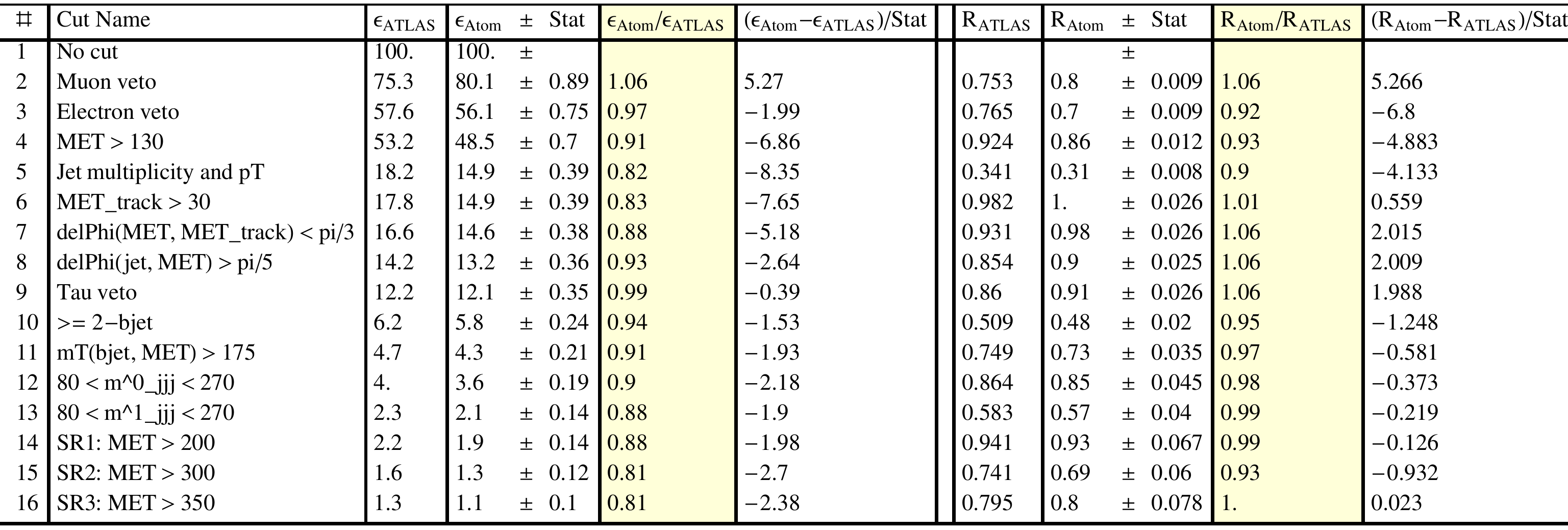}
    \vspace{-5mm}
\caption{ The same as in Fig.~\ref{fig:v_2013_024_tR} but with $\tilde t_L \tilde t_L^*$.
\label{fig:v_2013_024_tL} 
}
\end{center}
\end{figure}

\newpage
\subsection*{ATLAS\_CONF\_2013\_035}

\begin{itemize}
\item The events are generated using {\tt Herwig++~2.5.2} throughout this analysis. 
\end{itemize}

\begin{figure}[h!]
\begin{center}
  \includegraphics[width=1.0\textwidth]{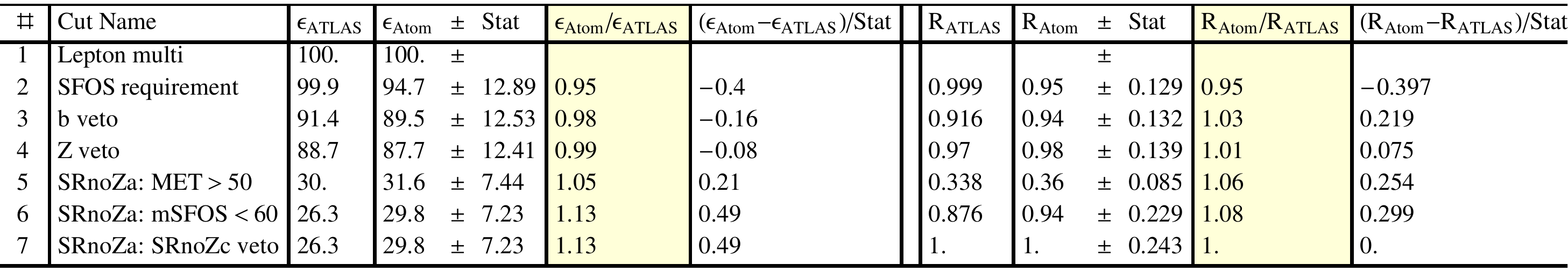} 
    \vspace{-5mm}
\caption{ ``noZa'' signal region in ATLAS\_CONF\_2013\_035.
$10^3$ events of $pp \to \chaone \ntwo$ process, followed by $\chaone \to \ell^\pm \nu \none$ and $\ntwo \to \ell^+ \ell^- \none$ both via an on-shell $\tilde \ell_L$, are used.  The masses are $m_\chaone = m_\ntwo = 192.5$~GeV, $m_{\tilde \ell_L} = 175$~GeV, $m_\none = 157.5$~GeV.
\label{fig:035noZa} 
}
\end{center}
\end{figure}

\begin{figure}[h!]
\begin{center}
  \includegraphics[width=1.0\textwidth]{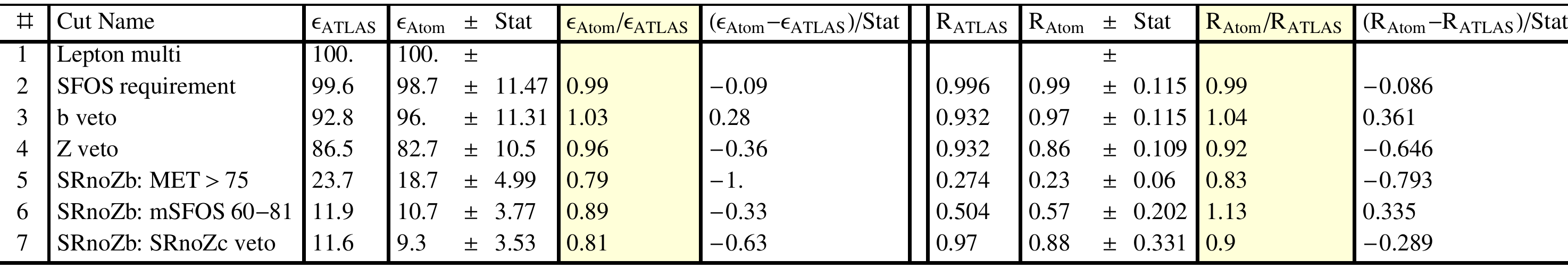} 
    \vspace{-5mm}
\caption{ ``noZb'' signal region in ATLAS\_CONF\_2013\_035.
$10^4$ events of $pp \to \chaone \ntwo \to W^\pm \none Z \none$ process are used.
The masses are: $m_\chaone = m_\ntwo = 150$~GeV, $m_\none = 75$~GeV. 
\label{fig:035noZb} 
}
\end{center}
\end{figure}

\begin{figure}[h!]
\begin{center}
  \includegraphics[width=1.0\textwidth]{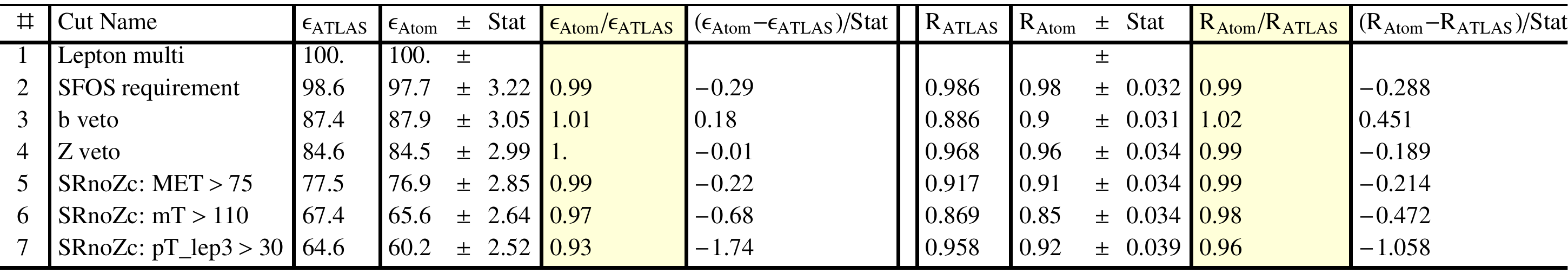}   
    \vspace{-5mm}
\caption{ ``noZc'' signal region in ATLAS\_CONF\_2013\_035.
$5 \cdot 10^3$ events of $pp \to \chaone \ntwo$ process, followed by $\chaone \to \ell^\pm \nu \none$ and $\ntwo \to \ell^+ \ell^- \none$ both via an on-shell $\tilde \ell_L$, are used.
The masses are: $m_\chaone = m_\ntwo = 500$~GeV, $m_{\tilde \ell_L} = 250$, $m_\none = 0$~GeV.  
\label{fig:035noZc} 
}
\end{center}
\end{figure}

\begin{figure}[h!]
\begin{center}
  \includegraphics[width=1.0\textwidth]{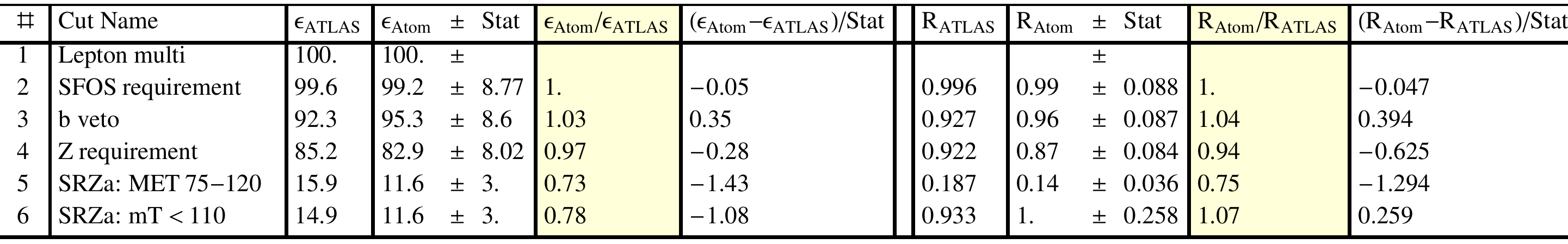}
     \vspace{-5mm}
\caption{ ``Za'' signal region in ATLAS\_CONF\_2013\_035.
$2 \cdot 10^4$ events of $pp \to \chaone \ntwo \to W^\pm \none Z \none$ process are used.
The masses are: $m_\chaone = m_\ntwo = 100$~GeV, $m_\none = 0$~GeV.
\label{fig:035Za} 
}
\end{center}
\end{figure}

\begin{figure}[h!]
\begin{center}
  \includegraphics[width=1.0\textwidth]{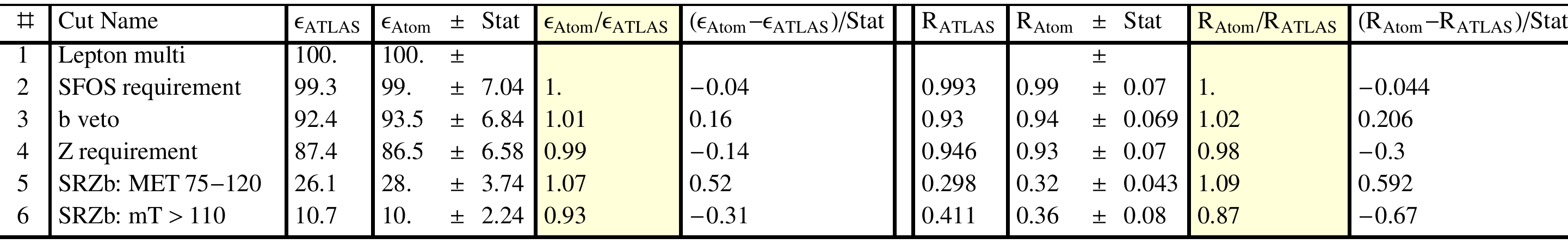} 
  \vspace{-5mm}  
\caption{ ``Zb'' signal region in ATLAS\_CONF\_2013\_035.
$3 \cdot 10^4$ events of $pp \to \chaone \ntwo \to W^\pm \none Z \none$ process are used.
The masses are: $m_\chaone = m_\ntwo = 150$~GeV, $m_\none = 0$~GeV.
\label{fig:035Zb} 
}
\end{center}
\end{figure}

\begin{figure}[h!]
\begin{center}
  \includegraphics[width=1.0\textwidth]{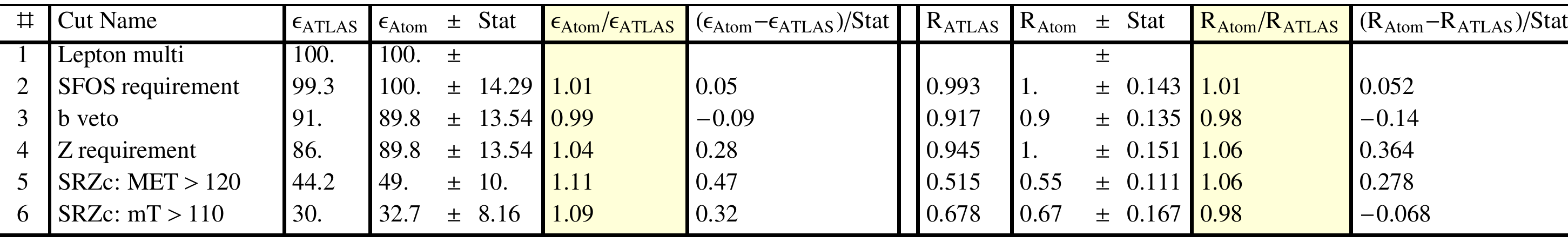}   
  \vspace{-5mm}
\caption{ ``Zc'' signal region in ATLAS\_CONF\_2013\_035.
$5 \cdot 10^3$ events of $pp \to \chaone \ntwo \to W^\pm \none Z \none$ process are used.
The masses are: $m_\chaone = m_\ntwo = 250$~GeV, $m_\none = 0$~GeV.
\label{fig:035Zc} 
}
\end{center}
\end{figure}

\newpage
~
\newpage

\subsection*{ATLAS\_CONF\_2013\_037}
\begin{itemize}
\item The events are generated using {\tt Herwig++~2.5.2} throughout this analysis. 
\end{itemize}
%
\begin{figure}[h!]
\begin{center}
  \includegraphics[width=1.0\textwidth]{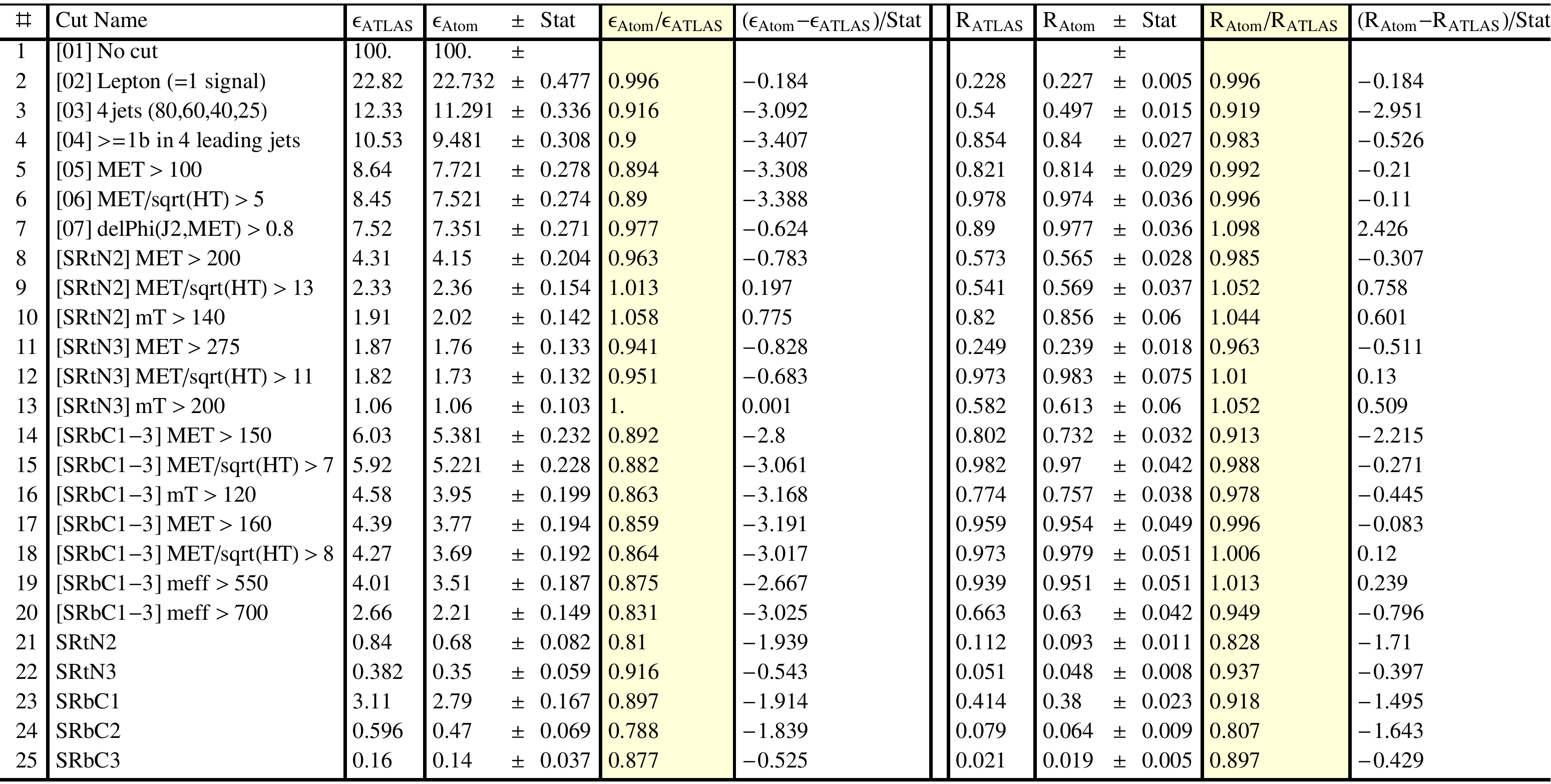} 
  \vspace{-7mm}  
  \caption{ATLAS\_CONF\_2013\_037 validation table. 
  $10^4$ events of $pp \to \tilde t_1 \tilde t_1^* \to t \none \bar t \none$ process are used with $m_{\tilde t_1} = 500$~GeV and $m_\none = 200$~GeV.
  In the cut stages 8, 14 and 21$-$25, $R_{\rm ATLAS/ATOM}$ is defied as the efficiency normalised by the efficiency at the stage 7.     
\label{fig:037_T1-500} 
}
\end{center}
\vspace{-5mm}
\end{figure}
\begin{figure}[h!]
\begin{center}
  \includegraphics[width=1.0\textwidth]{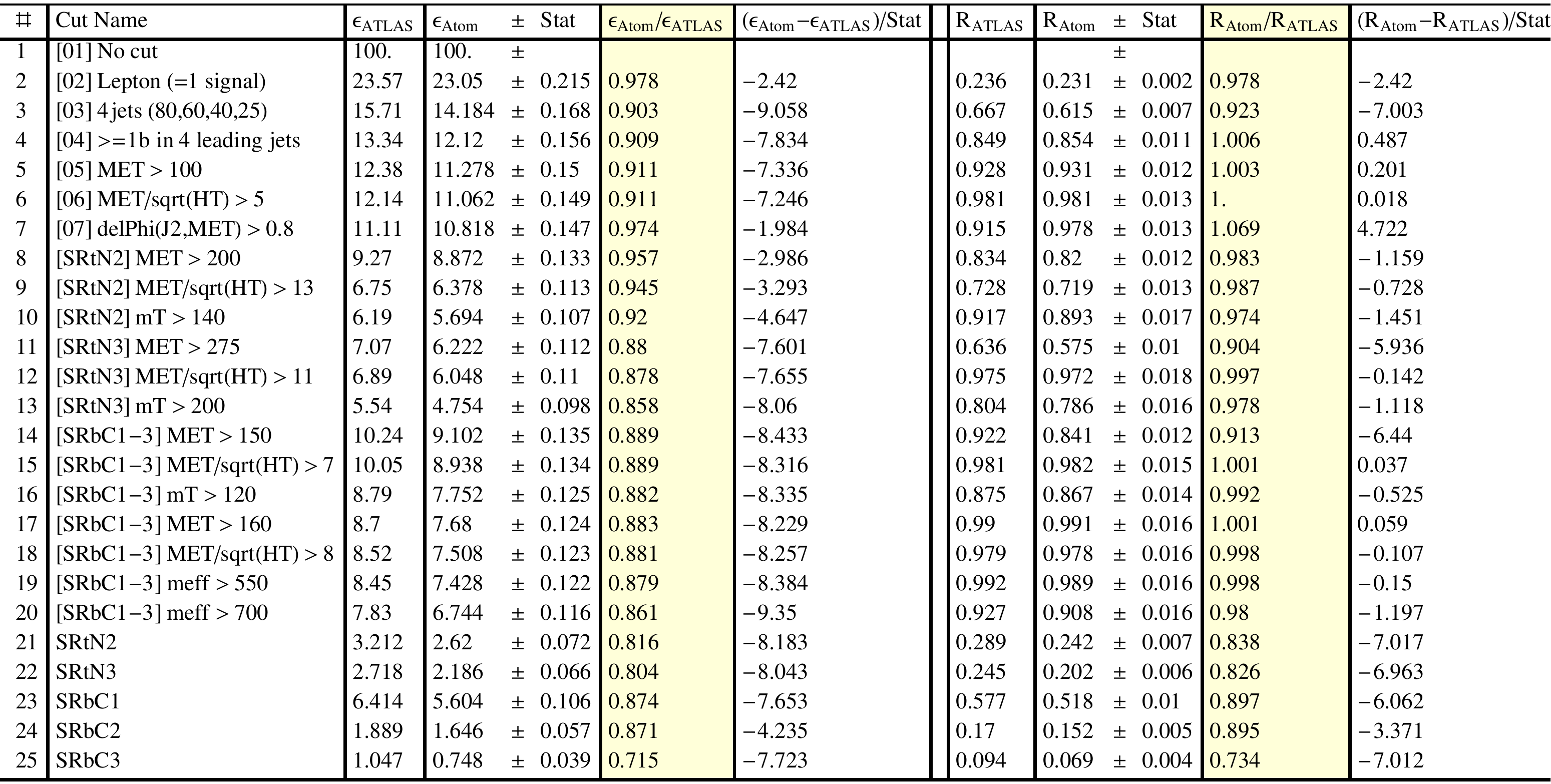} 
  \vspace{-7mm}    
  \caption{ The same as Fig.~\ref{fig:037_T1-500} but $5 \times 10^4$ events of $pp \to \tilde t_1 \tilde t_1^* \to t \none \bar t \none$ process with $m_{\tilde t_1} = 650$~GeV and $m_\none = 1$~GeV.
\label{fig:037_T1-650} 
}
\end{center}
\end{figure}

\newpage
\subsection*{ATLAS\_CONF\_2013\_047}

\begin{itemize}
\item The events are generated using {\tt MadGraph~5}~\cite{Alwall:2011uj} and {\tt Pythia~6}~\cite{Sjostrand:2006za} throughout this analysis. 
\item The MLM merging~\cite{Mangano:2006rw} is used with the shower-$k_T$ scheme implemented in {\tt MadGraph~5} and {\tt Pythia~6},
where we take ${\rm xqcut} = {\rm qcut} = m_{\rm SUSY}/4$ with $m_{\rm SUSY}$ being the mass of the heavier SUSY particles in the production. 
\end{itemize}
~
\vspace{-5mm}

\begin{figure}[h!]
\begin{center}
  \includegraphics[width=1.0\textwidth]{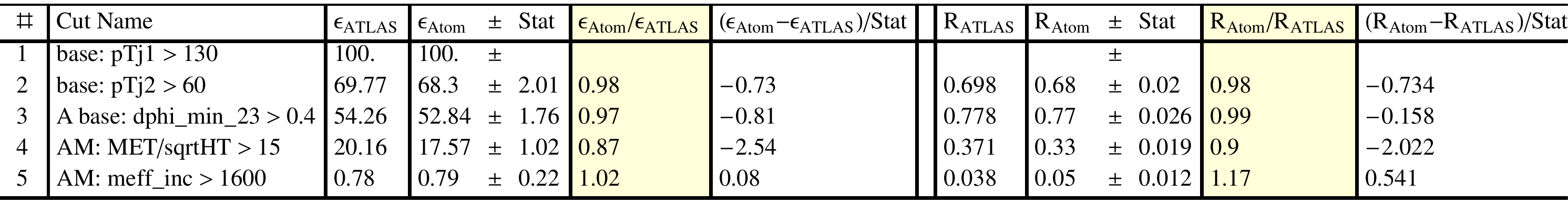} 
     \vspace{-7mm}  
  \caption{ 
  ``A medium'' signal region in ATLAS\_CONF\_2013\_047.
  $2 \cdot 10^4$ events of $pp \to \tilde q \tilde q \to q \none q \none$ process are used.
  The masses are: $m_{\tilde q} = 450$~GeV, $m_\none = 400$~GeV.
\label{fig:047AM450} 
}
\end{center}
\end{figure}

\begin{figure}[h!]
\begin{center}
  \includegraphics[width=1.0\textwidth]{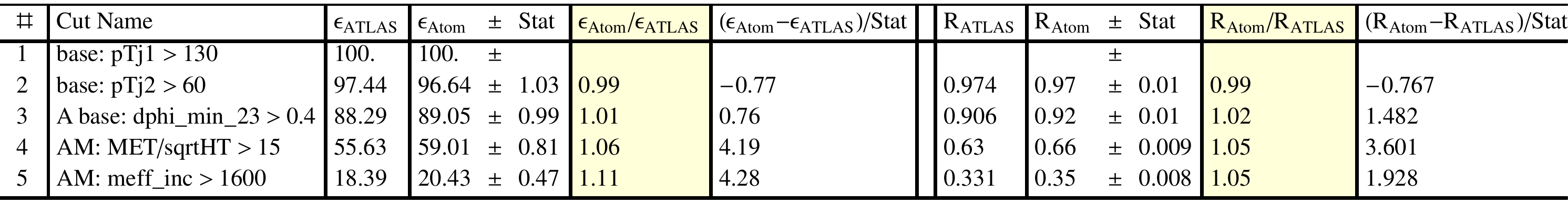} 
     \vspace{-7mm}    
  \caption{   ``A medium'' signal region in ATLAS\_CONF\_2013\_047.
  $10^4$ events of $pp \to \tilde q \tilde q \to q \none q \none$ process are used.
  The masses are: $m_{\tilde q} = 850$~GeV, $m_\none = 100$~GeV.
\label{fig:047AM850} 
}
\end{center}
\end{figure}

\begin{figure}[h!]
\begin{center}
  \includegraphics[width=1.0\textwidth]{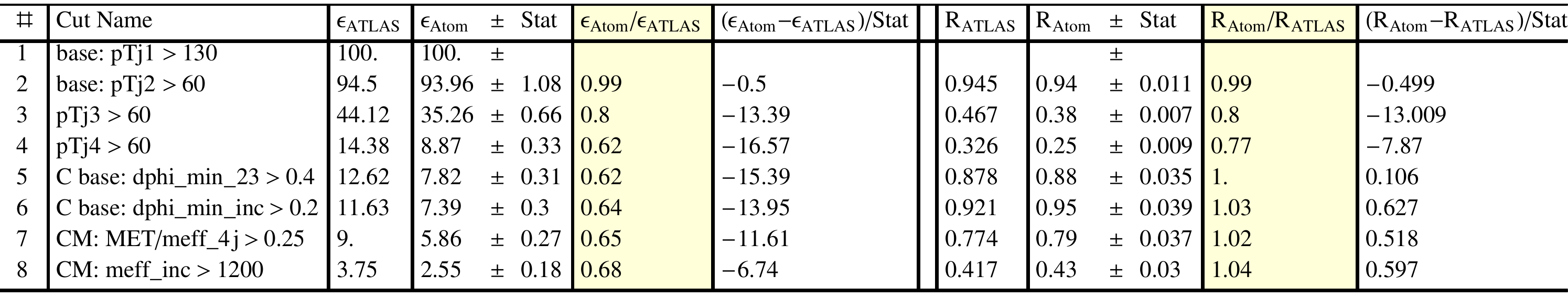} 
     \vspace{-7mm}    
  \caption{   ``C medium'' signal region in ATLAS\_CONF\_2013\_047.
  $10^4$ events of $pp \to \tilde q \tilde q \to q \none q \none$ process are used.
  The masses are: $m_{\tilde q} = 662$~GeV, $m_\none = 287$~GeV.
\label{fig:047CM} 
}
\end{center}
\end{figure}

\begin{figure}[h!]
\begin{center}
  \includegraphics[width=1.0\textwidth]{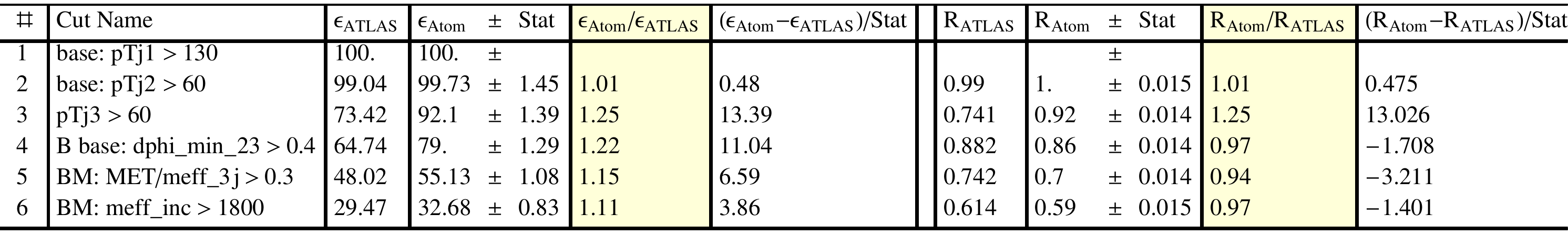} 
     \vspace{-7mm}    
  \caption{   ``B medium'' signal region in ATLAS\_CONF\_2013\_047.
  $5 \cdot 10^3$ events of $pp \to \tilde g \tilde q$ process, followed by $\tilde g \to q \bar q \none$ and $\tilde q \to q \none$, are used.
  The masses are: $m_\gluino = 1425$~GeV, $m_{\tilde q} = 1368$~GeV and $m_\none = 525$~GeV.
}
\end{center}
\end{figure}

\begin{figure}[h!]
\begin{center}
  \includegraphics[width=1.0\textwidth]{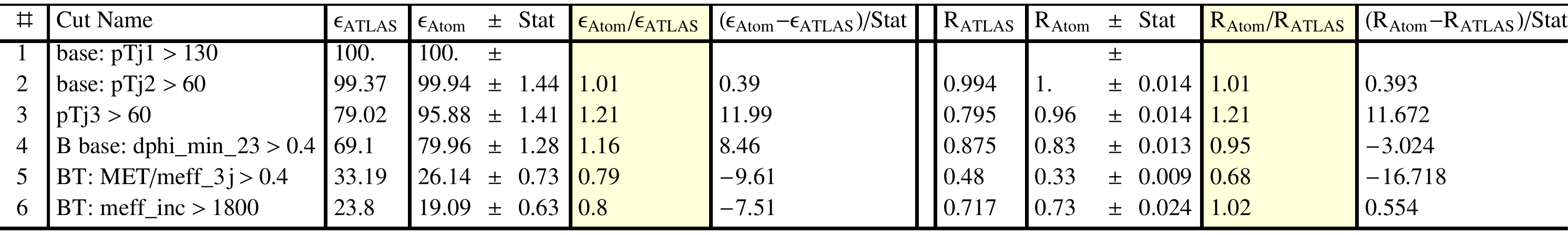} 
     \vspace{-7mm}    
  \caption{ ``B tight'' signal region in ATLAS\_CONF\_2013\_047.
  $5 \cdot 10^3$ events of $pp \to \tilde g \tilde q$ process, followed by $\tilde g \to q \bar q \none$ and $\tilde q \to q \none$, are used.
  The masses are: $m_\gluino = 1612$~GeV, $m_{\tilde q} = 1548$~GeV and $m_\none = 37$~GeV.
\label{fig:047BT} 
}
\end{center}
\end{figure}

\begin{figure}[h!]
\begin{center}
  \includegraphics[width=1.0\textwidth]{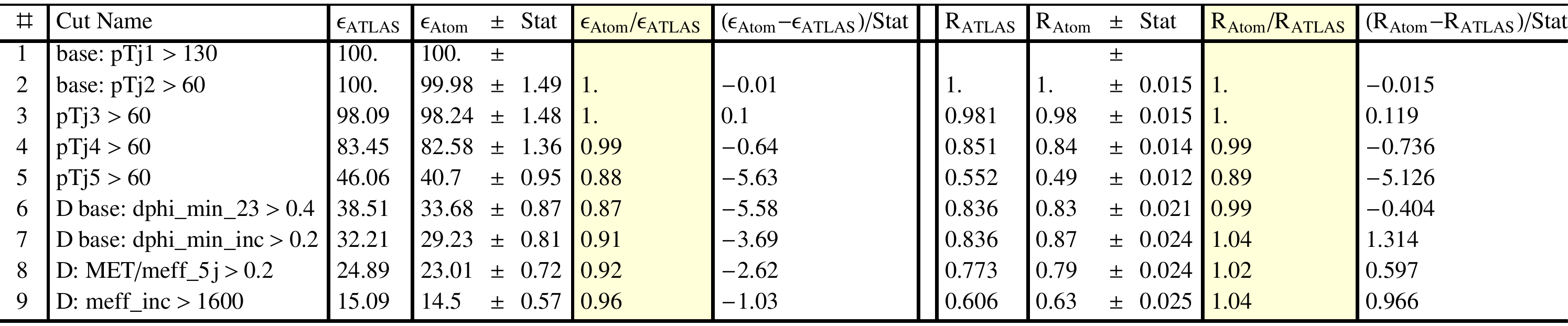} 
     \vspace{-7mm}    
  \caption{ ``D'' signal region in ATLAS\_CONF\_2013\_047.
  $5 \cdot 10^3$ events of $pp \to \tilde g \tilde q \to qq \none q \none$ process are used.
  The masses are: $m_\gluino = 1162$~GeV and $m_\none = 337$~GeV.
\label{fig:047Ddirect} 
}
\end{center}
\end{figure}

\begin{figure}[h!]
\begin{center}
  \includegraphics[width=1.0\textwidth]{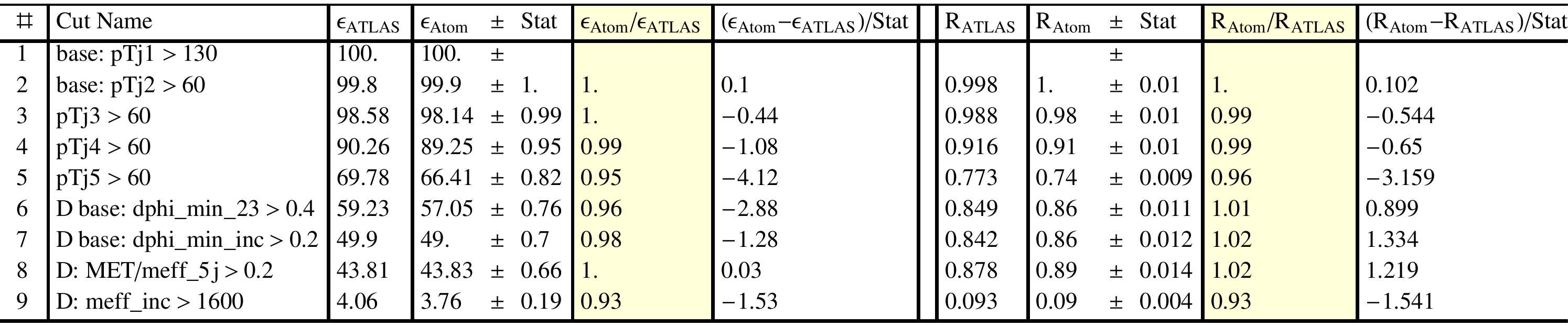} 
       \vspace{-7mm}  
  \caption{ ``D'' signal region in ATLAS\_CONF\_2013\_047.
  $2 \cdot 10^4$ events of $pp \to \tilde g \tilde g$ followed by $\gluino \to qq \chaone \to qq W^\pm \none$ are used.
  The masses are: $m_\gluino = 1065$~GeV, $m_\chaone = 785$~GeV and $m_\none = 505$~GeV. 
\label{fig:D1065} 
}
\end{center}
\end{figure}

\begin{figure}[h!]
\begin{center}
  \includegraphics[width=1.0\textwidth]{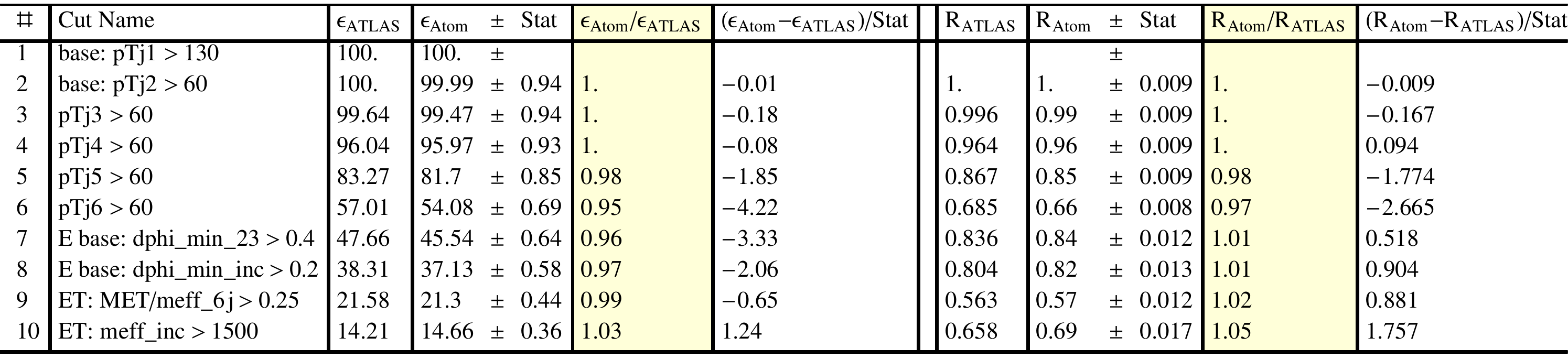} 
     \vspace{-7mm}    
  \caption{  ``E tight'' signal region in ATLAS\_CONF\_2013\_047.
  $2 \cdot 10^4$ events of $pp \to \tilde g \tilde g$ followed by $\gluino \to qq \chaone \to qq W^\pm \none$ are used.
  The masses are: $m_\gluino = 1265$~GeV, $m_\chaone = 865$~GeV and $m_\none = 465$~GeV.
\label{fig:ET1265} 
}
\end{center}
\end{figure}

\newpage
~
\newpage
\subsection*{ATLAS\_CONF\_2013\_048}

\begin{itemize}
\item The events are generated using {\tt MadGraph~5} and {\tt Pythia~6}. 
\end{itemize}
%

\begin{figure}[h!]
\begin{center}
  \vspace{-3mm}
  \subfigure[]{\includegraphics[width=1.\textwidth]{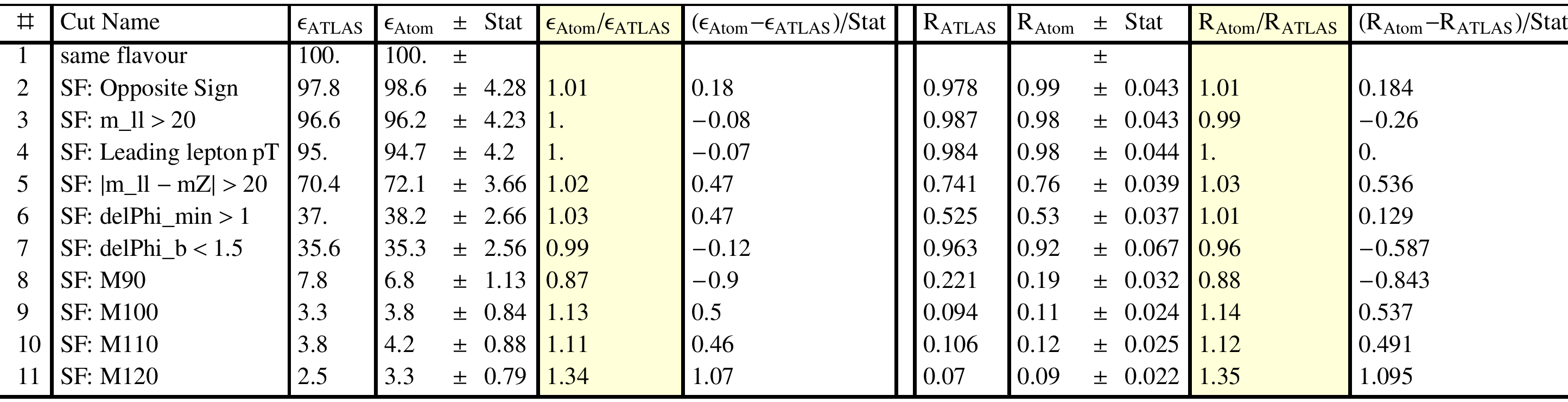}\label{fig:048SF}}
  \vspace{-3mm}
  \subfigure[]{\includegraphics[width=1.\textwidth]{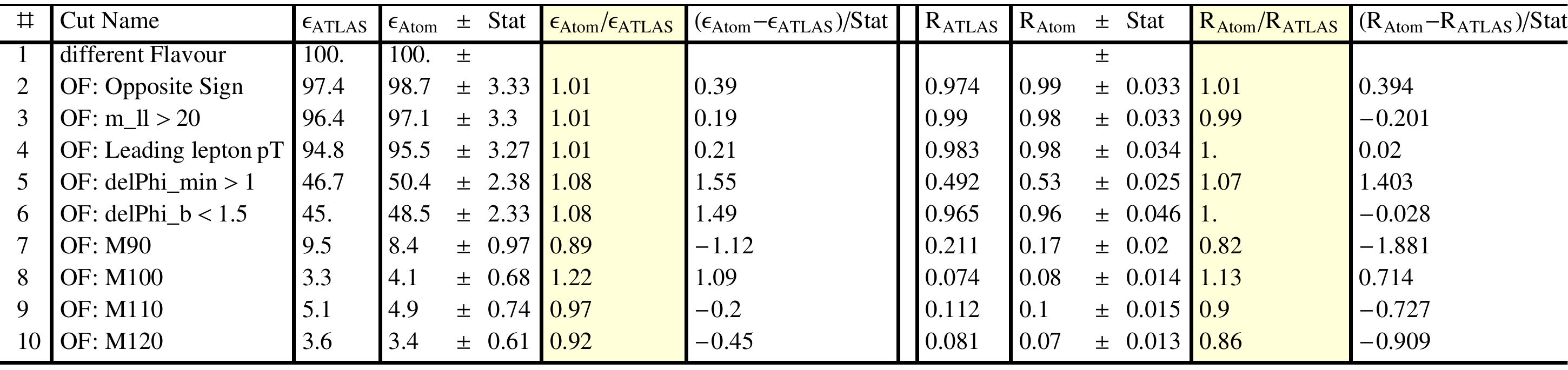}\label{fig:048OF}} 
     \vspace{-1mm}      
  \caption{ The signal regions in the same (a) and opposite (b) flavour channels in ATLAS\_CONF\_2013\_048.
  $3 \cdot 10^4$ events of $pp \to \tilde t_1 \tilde t_1^*$ followed by $\tilde t_1 \to b \chaone \to b W^+ \none$ are used.
  The masses are: $m_{\tilde t_1} = 400$~GeV, $m_\chaone = 250$~GeV and $m_\none = 1$~GeV.
\label{fig:048} 
}
\end{center}
\end{figure}

\vspace{-3mm}

\subsection*{ATLAS\_CONF\_2013\_049}

\begin{itemize}
\item The events are generated using {\tt Herwig++~2.5.2} throughout this analysis. 
\end{itemize}
~
\vspace{-6mm}

\begin{figure}[h!]
\begin{center}
  \includegraphics[width=1.0\textwidth]{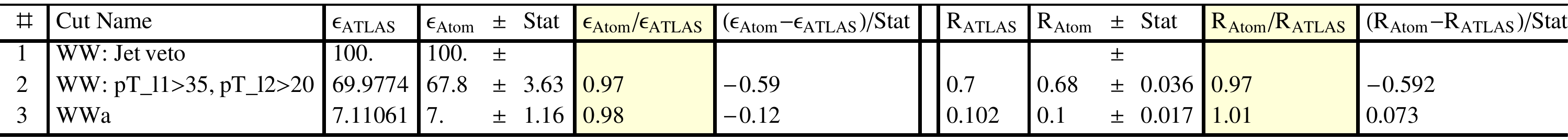} 
  \includegraphics[width=1.0\textwidth]{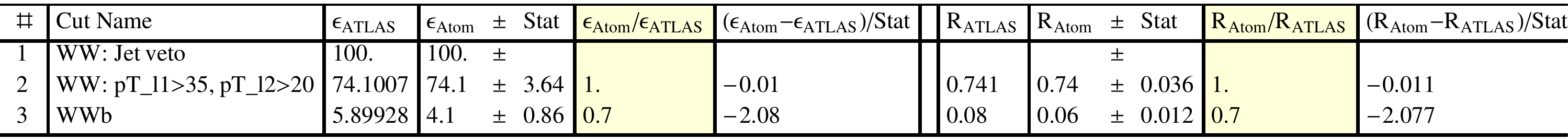}   
  \includegraphics[width=1.0\textwidth]{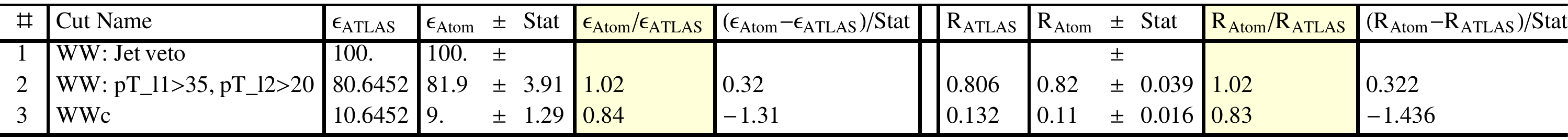}     
     \vspace{-6mm}      
  \caption{ ``Wa'' (top), ``Wb'' (middle) and ``Wc'' (bottom) signal regions in ATLAS\_CONF\_2013\_049.
  $5 \cdot 10^4$ events of $pp \to \chaonep \chaonem \to W^+ \none W^- \none$ process are used.
  The masses are: 
  $(m_\chaone, m_\none) = (100, 0)$~GeV for Wa, 
  $(m_\chaone, m_\none) = (140, 20)$~GeV for Wb and 
  $(m_\chaone, m_\none) = (200, 0)$~GeV for Wc.
  \label{fig:049WW} 
}
\end{center}
\end{figure}

\begin{figure}[h!]
\begin{center}
  \includegraphics[width=1.0\textwidth]{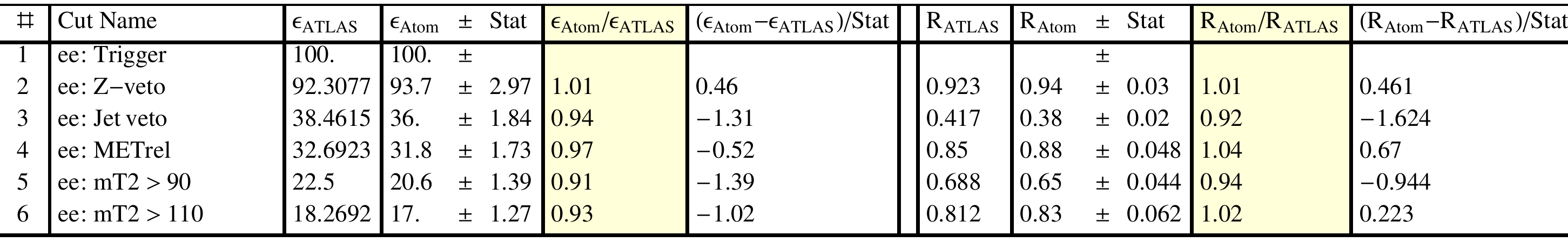} 
  \includegraphics[width=1.0\textwidth]{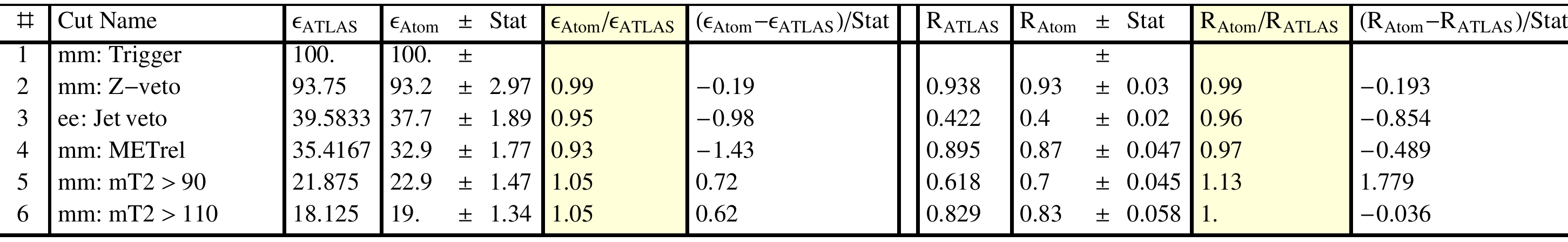}   
  \includegraphics[width=1.0\textwidth]{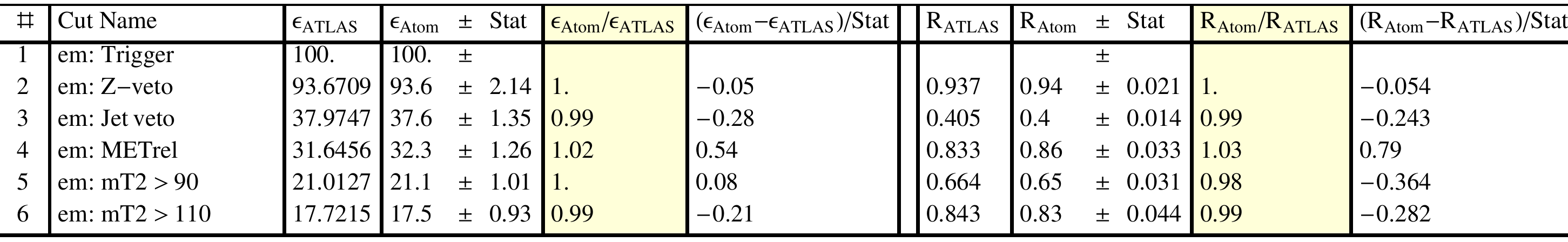}     
     \vspace{-5mm}      
  \caption{ ``mT2:90'' and ``mT2:100'' signal regions in the $ee$ (top), $\mu \mu$ (middle) and $e \mu$ (bottom) channels
  in ATLAS\_CONF\_2013\_049.
  $10^4$ events of $pp \to \chaonep \chaonem$ process, 
  followed by $\chaone \to \ell^\pm_i \nu \none$ via an on-shell $\tilde \ell_i$, are used.
  The masses are: $m_\chaone = 350$~GeV, $m_{\tilde \ell} = 175$~GeV and $m_\none = 0$~GeV.
\label{fig:049CLN350} 
}
\end{center}
\end{figure}

\begin{figure}[h!]
\begin{center}
  \includegraphics[width=1.0\textwidth]{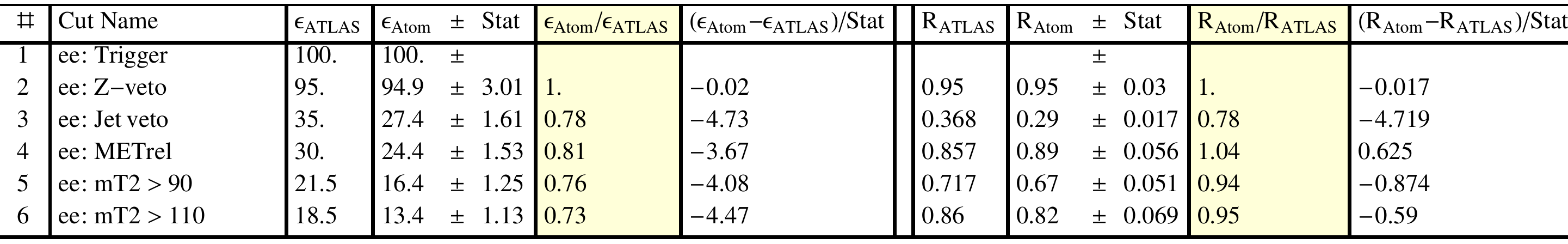} 
  \includegraphics[width=1.0\textwidth]{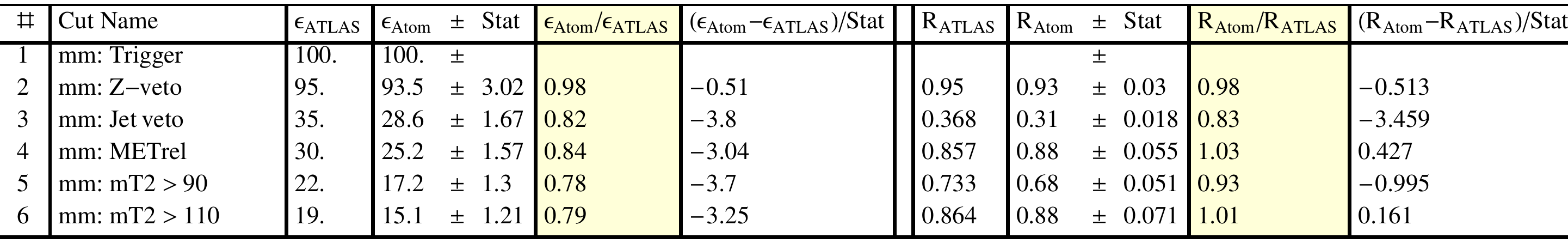}   
  \includegraphics[width=1.0\textwidth]{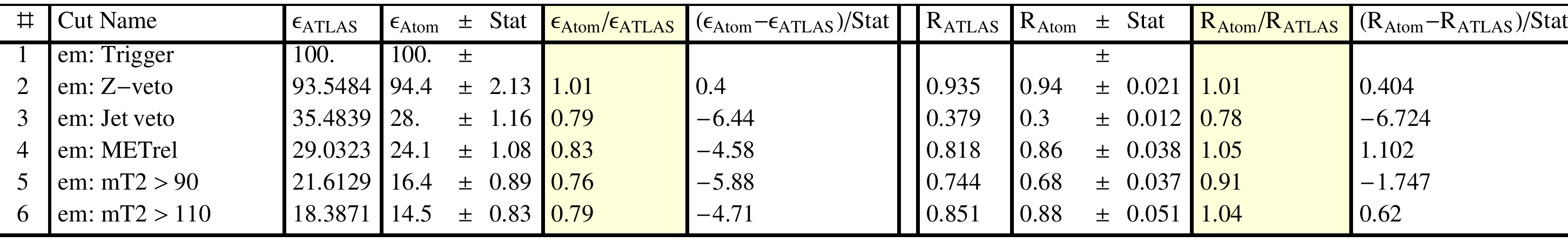}     
     \vspace{-5mm}      
  \caption{ The same as Fig.~\ref{fig:049CLN350} but
   $m_\chaone = 425$~GeV, $m_{\tilde \ell} = 250$~GeV and $m_\none = 75$~GeV. 
\label{fig:049CLN400} 
}
\end{center}
\end{figure}

\begin{figure}[h!]
\begin{center}
  \includegraphics[width=1.0\textwidth]{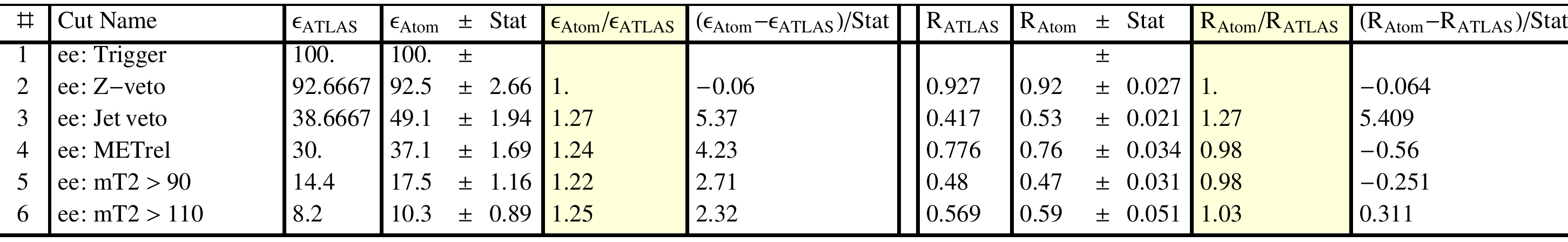} 
  \includegraphics[width=1.0\textwidth]{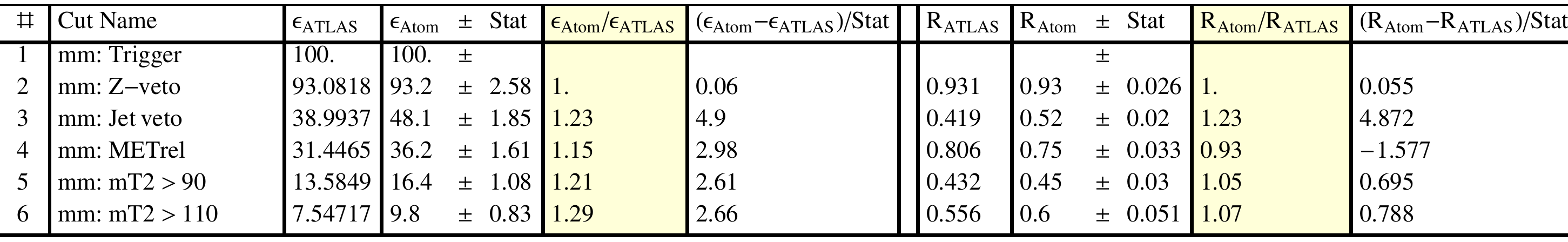}   
     \vspace{-5mm}      
  \caption{``mT2:90'' and ``mT2:100'' signal regions in the $ee$ (top) and $\mu \mu$ (middle) channels
  in ATLAS\_CONF\_2013\_049.
  $2 \cdot 10^3$ events of $pp \to \tilde e^+ \tilde e^- \to e^+ \none e^- \none$ and 
  $pp \to \tilde \mu^+ \tilde \mu^- \to \mu^+ \none \mu^- \none$ processes are used for $ee$ and $\mu \mu$ channels, respectively.
  The masses are: $m_{\tilde \ell} = 191$~GeV and $m_\none = 90$~GeV.
\label{fig:049LN191} 
}
\end{center}
\end{figure}

\begin{figure}[h!]
\begin{center}
  \includegraphics[width=1.0\textwidth]{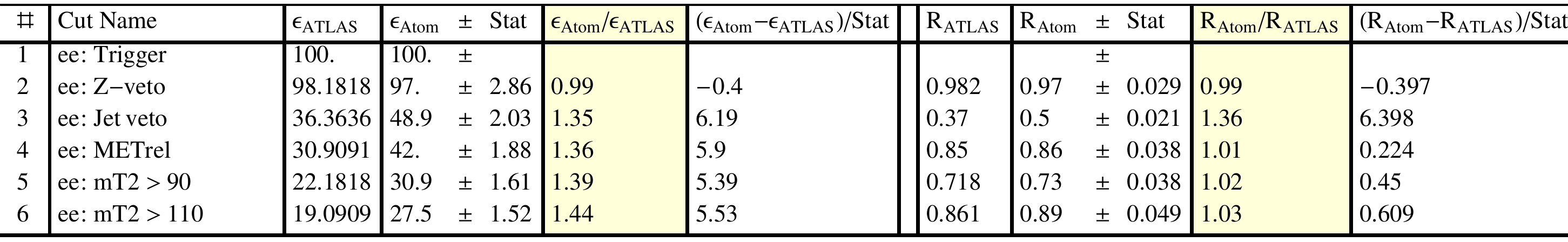} 
  \includegraphics[width=1.0\textwidth]{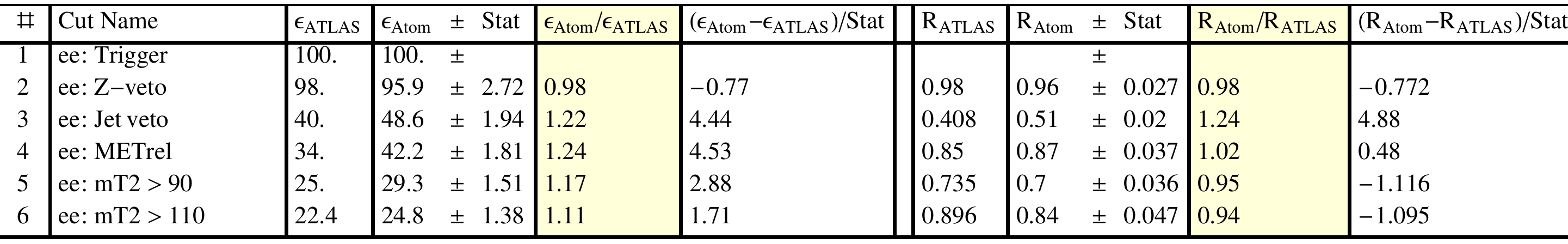}   
     \vspace{-5mm}      
  \caption{ The same as Fig.~\ref{fig:049LN250} but
   $m_{\tilde \ell} = 250$~GeV and $m_\none = 10$~GeV. 
   \label{fig:049LN250} 
}
\end{center}
\end{figure}

~
\newpage
~
\newpage
~
\newpage

\subsection*{ATLAS\_CONF\_2013\_053}

\begin{itemize}
\item The events are generated using {\tt MadGraph~5} and {\tt Pythia~6}. 
\end{itemize}
~
\vspace{-7mm}

\begin{figure}[h!]
\begin{center}
  \includegraphics[width=0.5\textwidth]{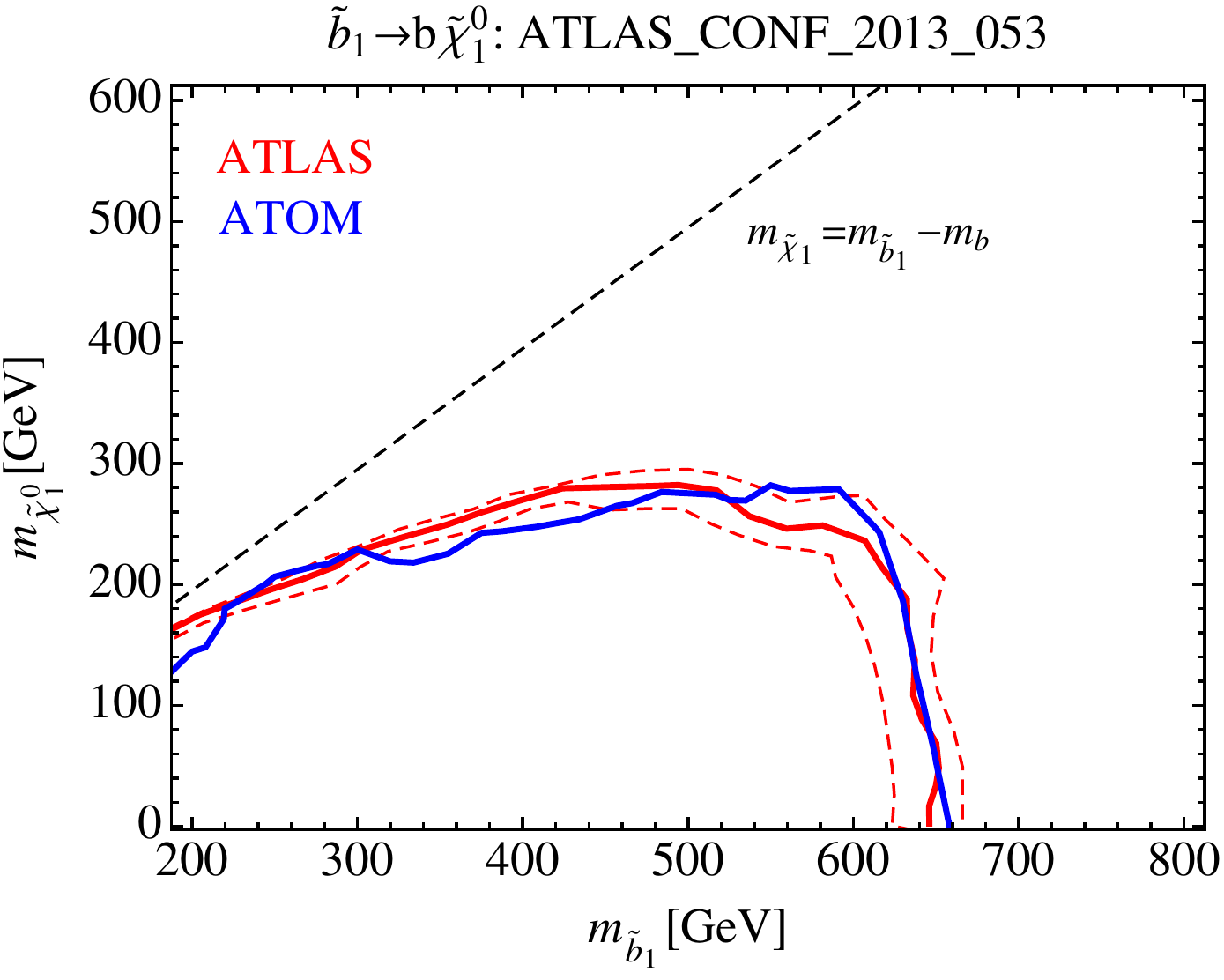}
     \vspace{-3mm}      
\caption{ The exclusion curves in the $\tilde b_1$-$\none$ simplified model parameter space ($\tilde b_1 \to b \none$).
The red and blue curves are for ATLAS and {\tt ATOM}, respectively.
The red dashed curves show the 1-$\sigma$ error band of the ATLAS exclusion curve.
\label{fig:053} 
}
\end{center}
\end{figure}

\vspace{-5mm}
\subsection*{ATLAS\_CONF\_2013\_054}

\begin{itemize}
\item The events are generated using {\tt MadGraph~5} and {\tt Pythia~6}.  
The MLM merging is used in {\tt MadGraph~5} and {\tt Pythia~6} with ${\rm xqcut} = {\rm qcut} = m_{\gluino}/4$.
\end{itemize}
~
\vspace{-5mm}

\begin{figure}[h!]
\begin{center}
  \includegraphics[width=0.5\textwidth]{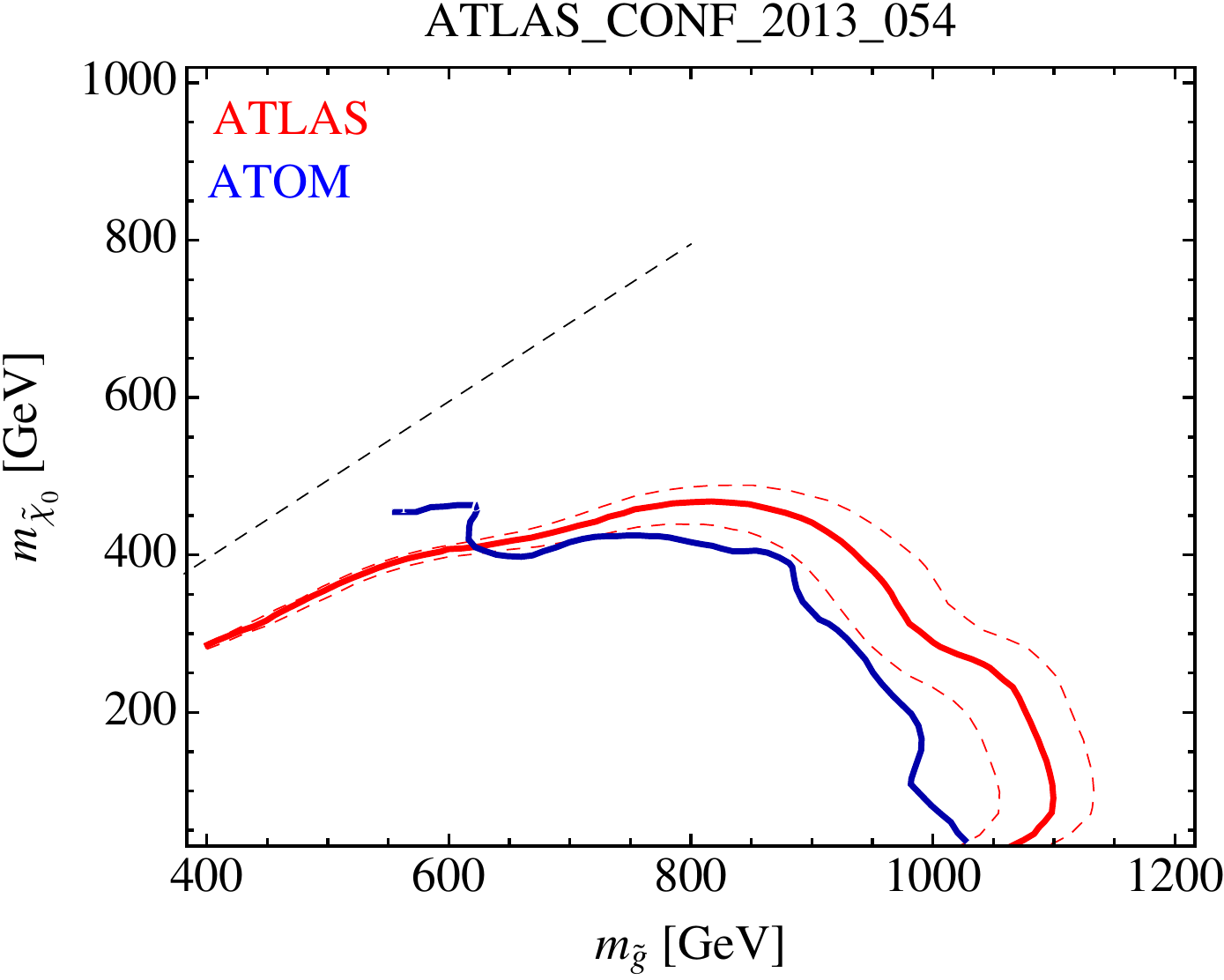}
     \vspace{-3mm}      
\caption{ The exclusion curves in the $\tilde g$-$\chaone$-$\none$ simplified model parameter space 
($\tilde g \to qq \chaone \to qq W^\pm \none$).
The chargino mass fixed at $(m_\gluino  - m_\none)/2$.
The red and blue curves are for ATLAS and {\tt ATOM}, respectively.
The red dashed curves show the 1-$\sigma$ error band of the ATLAS exclusion curve.
\label{fig:054} 
}
\end{center}
\end{figure}

\newpage
\subsection*{ATLAS\_CONF\_2013\_061}

\begin{figure}[h!]
\begin{center}
  \includegraphics[width=1.0\textwidth]{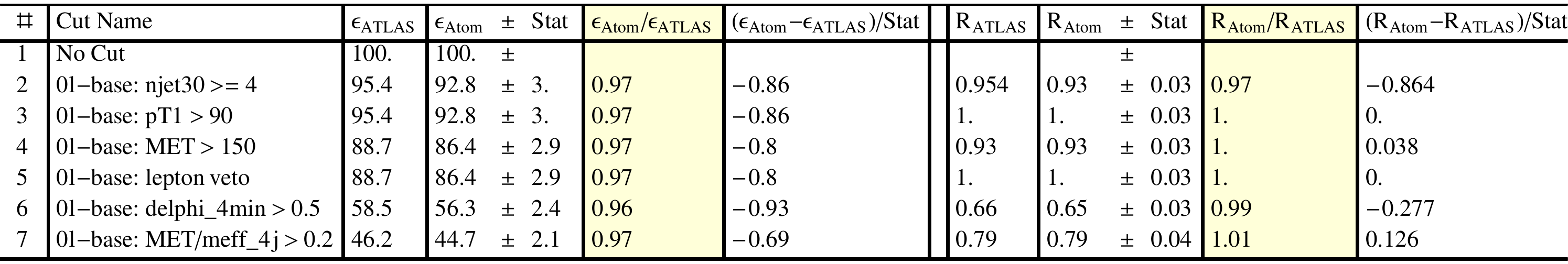} 
  \includegraphics[width=1.0\textwidth]{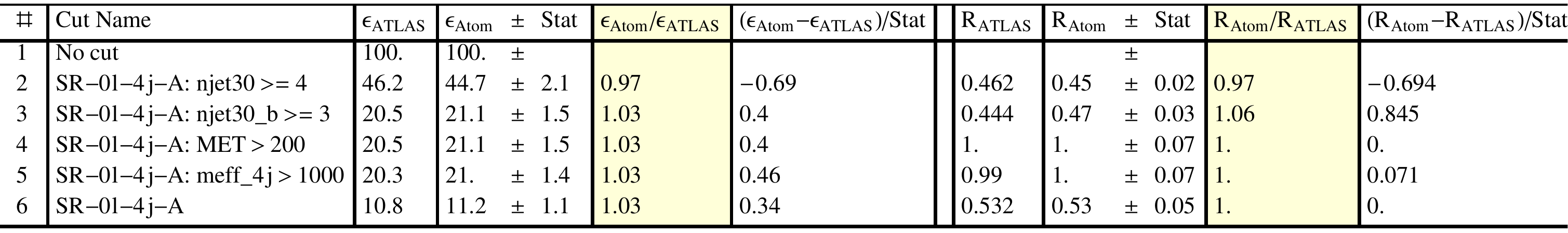} 
  \includegraphics[width=1.0\textwidth]{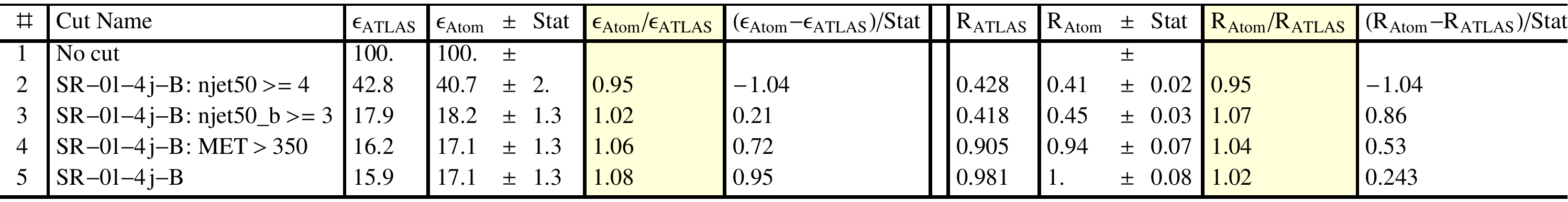} 
  \includegraphics[width=1.0\textwidth]{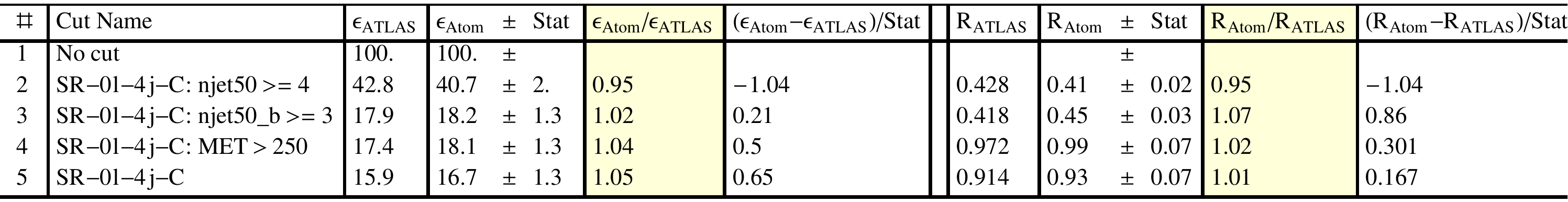}   
  \vspace{-7mm}
  \caption{ The 0-lepton + 4-jet signal regions in ATLAS\_CONF\_2013\_061.
  $10^3$ events of $pp \to \gluino \gluino \to b \bar b \none b \bar b \none$
  process generated by {\tt MadGraph~5} are used.
  The masses are: $m_\gluino = 1300$~GeV and $m_\none = 100$~GeV. 
\label{fig:061-0L4jbb} 
}
\end{center}
\end{figure}

\vspace{-5mm}

\begin{figure}[h!]
\begin{center}
  \includegraphics[width=1.0\textwidth]{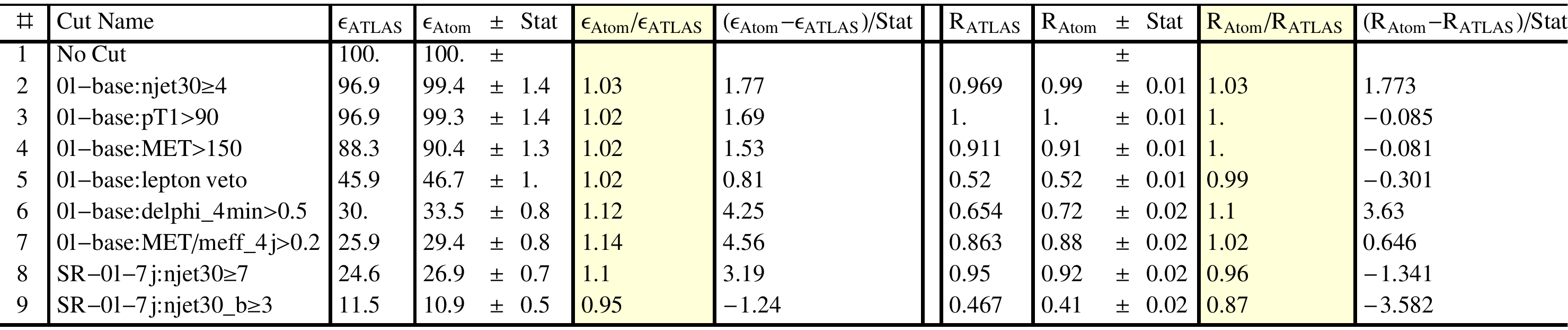} 
  \includegraphics[width=1.0\textwidth]{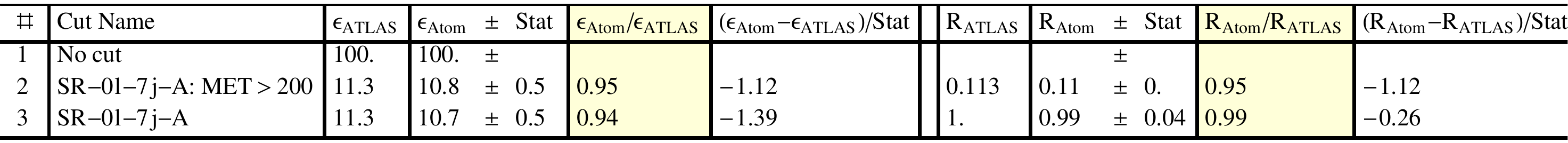} 
  \includegraphics[width=1.0\textwidth]{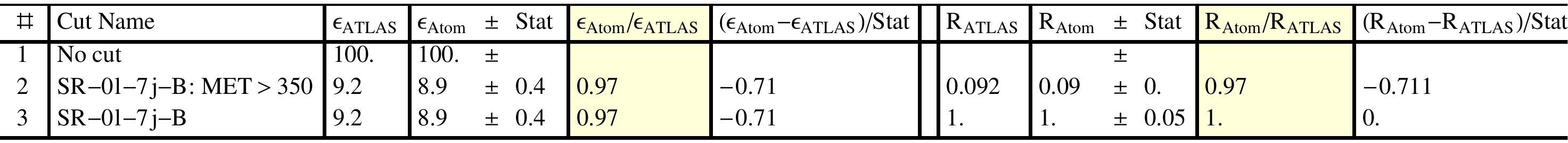} 
  \includegraphics[width=1.0\textwidth]{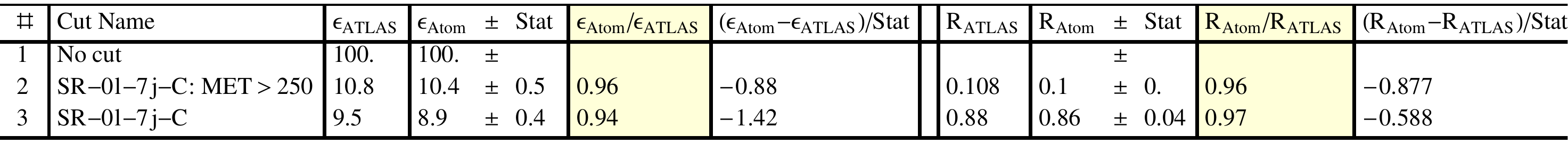}   
  \vspace{-7mm}  
  \caption{ The 0-lepton + 7-jet signal regions in ATLAS\_CONF\_2013\_061.
  $5 \cdot 10^3$ events of $pp \to \gluino \gluino \to t \bar t \none t \bar t \none$
  process generated by {\tt Herwig++~2.5.2} are used.
  The masses are: $m_\gluino = 1300$~GeV and $m_\none = 100$~GeV. 
\label{fig:061-0L7jtt} 
}
\end{center}
\end{figure}

\begin{figure}[h!]
\begin{center}
  \includegraphics[width=1.0\textwidth]{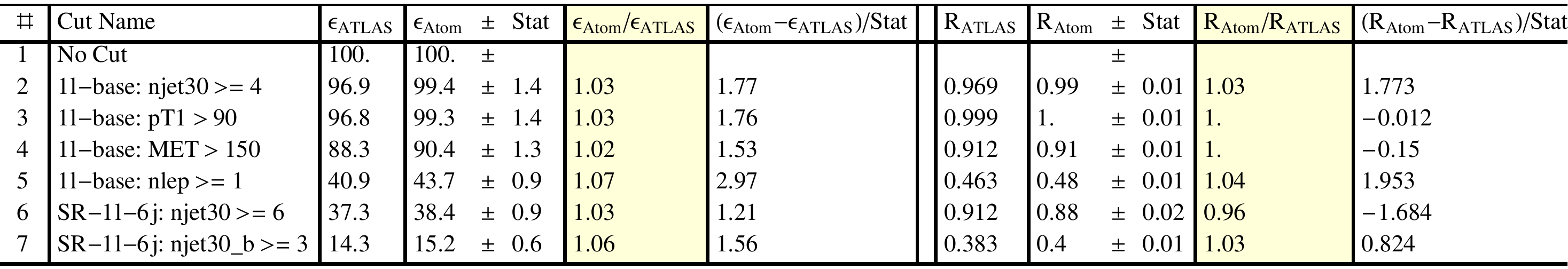} 
  \includegraphics[width=1.0\textwidth]{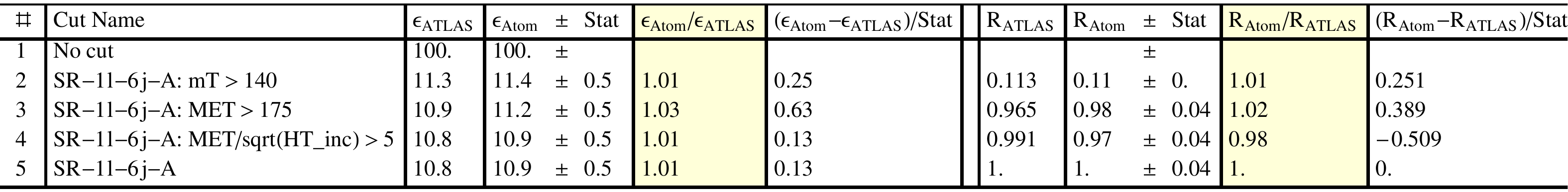} 
  \includegraphics[width=1.0\textwidth]{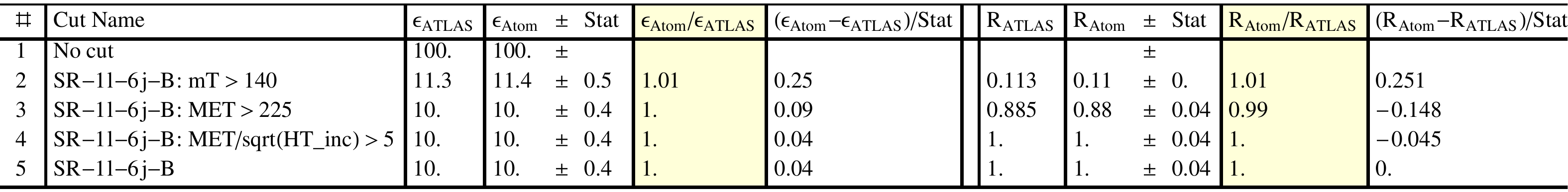} 
  \includegraphics[width=1.0\textwidth]{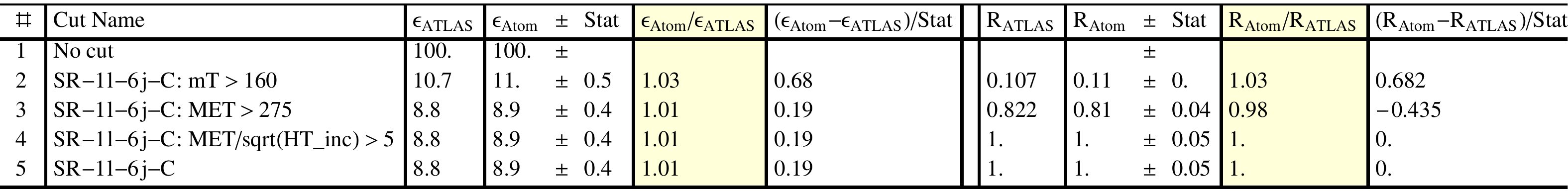}   
  \vspace{-7mm}  
  \caption{ The 1-lepton + 6-jet signal regions in ATLAS\_CONF\_2013\_061.
  $5 \cdot 10^3$ events of $pp \to \gluino \gluino \to t \bar t \none t \bar t \none$
  process generated by {\tt Herwig++~2.5.2} are used.
  The masses are: $m_\gluino = 1300$~GeV and $m_\none = 100$~GeV. 
\label{fig:061-1L} 
}
\end{center}
\end{figure}

\vspace{-10mm}
\subsection*{ATLAS\_CONF\_2013\_093}

\begin{itemize}
\item The events are generated using {\tt Herwig++~2.5.2}.  
\end{itemize}
\vspace{-3mm}

\begin{figure}[h!]
\begin{center}
  \includegraphics[width=1.0\textwidth]{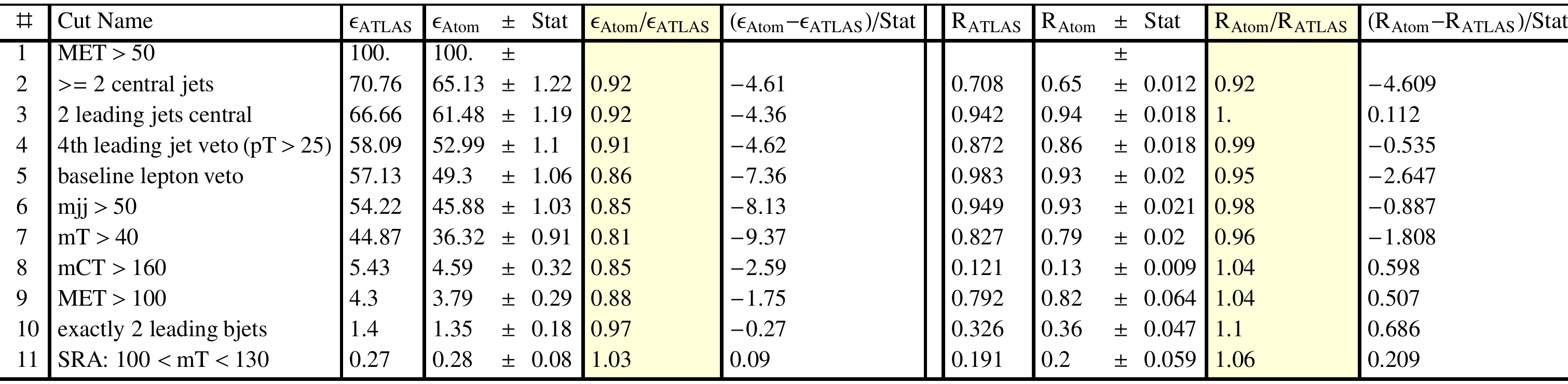} 
  \includegraphics[width=1.0\textwidth]{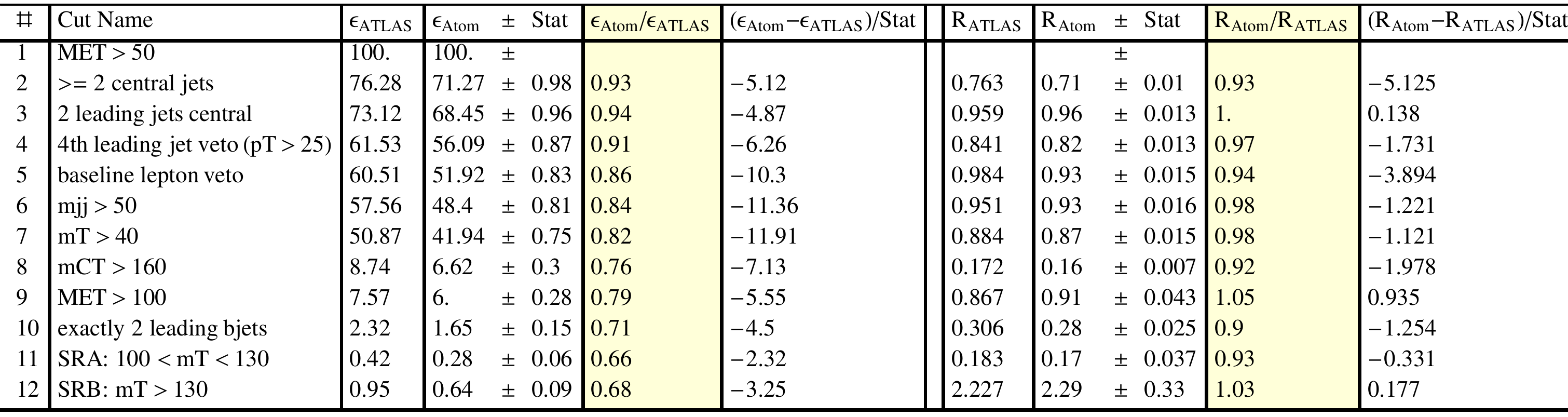} 
  \vspace{-7mm}  
  \caption{ The signal regions in ATLAS\_CONF\_2013\_093.
  $5 \cdot 10^4$ events of $pp \to \chaone \ntwo \to W^\pm \none h^0 \none$ process are used.
  The masses are: $m_\chaone = m_\ntwo = 130 (225)$~GeV, $m_\none = 0$~GeV for the top (bottom) table. 
\label{fig:093} 
}
\end{center}
\end{figure}

\providecommand{\href}[2]{#2}\begingroup\raggedright

\end{document}